\documentclass{article}

     \usepackage[final,nonatbib]{neurips_2018}

\usepackage[utf8]{inputenc} 
\usepackage[T1]{fontenc} 
\usepackage{hyperref} 
\usepackage{url} 
\usepackage{booktabs} 
\usepackage{amsfonts} 
\usepackage{nicefrac} 
\usepackage{microtype} 
\usepackage{cite}

\usepackage{enumitem}
\usepackage{wrapfig}
\usepackage{caption}
\usepackage{subcaption}
\usepackage{amsmath, amsthm}

\usepackage{graphicx}
\usepackage{bm}

\usepackage[algo2e]{algorithm2e}
\usepackage{multirow}
\usepackage{makecell}

\usepackage{stackengine}

\makeatletter
\newcommand{\distas}[1]{\mathbin{\overset{#1}{\kern\z@\sim}}}%

\newtheorem{theorem}{Theorem}
\newtheorem{definition}{Definition}

\newtheorem{cor}{Corollary}

\newtheorem{property}{Property}

\newcommand{\beq}{\vspace{0mm}\begin{equation}}
\newcommand{\eeq}{\vspace{0mm}\end{equation}}
\newcommand{\beqs}{\vspace{0mm}\begin{eqnarray}}
\newcommand{\eeqs}{\vspace{0mm}\end{eqnarray}}
\newcommand{\barr}{\begin{array}}
\newcommand{\earr}{\end{array}}

\newcommand{\Imat}{{\bf I}}

\newcommand{\xv}{\boldsymbol{x}}

\newcommand{\betav}[0]{{\boldsymbol{\beta}}}

\newcommand{\E}{\mathbb{E}}

\newcommand{\given }{\,|\,}

\usepackage{changes}

\usepackage{wrapfig,lipsum,booktabs}

\usepackage{siunitx}

\usepackage{color,soul}

\title{
Nonparametric Bayesian Lomax delegate racing \\ for survival analysis with competing risks
}
\author{
 Quan Zhang\\
McCombs School of Business\\
 The University of Texas at Austin\\
 Austin, TX 78712 \\
 \texttt{quan.zhang@mccombs.utexas.edu} \\
 \And
 Mingyuan Zhou\\
McCombs School of Business\\
 The University of Texas at Austin\\
 Austin, TX 78712 \\
 \texttt{mingyuan.zhou@mccombs.utexas.edu} \\
}
\begin{document}

\maketitle

\begin{abstract}
We propose Lomax delegate racing (LDR) to explicitly model the mechanism of survival under competing risks and to interpret how the covariates accelerate or decelerate the time to event. LDR explains non-monotonic covariate effects by racing a potentially infinite number of sub-risks, and consequently relaxes the ubiquitous proportional-hazards assumption which may be too restrictive. Moreover, LDR is naturally able to model not only censoring, but also missing event times or event types. For inference, we develop a Gibbs sampler under data augmentation for moderately sized data, along with a stochastic gradient descent maximum a posteriori inference algorithm for big data applications. Illustrative experiments are provided on both synthetic and real datasets, and comparison with various benchmark algorithms for survival analysis with competing risks demonstrates distinguished performance of LDR. 
\end{abstract}

\section{Introduction}
In survival analysis, one can often use nonparametric approaches to flexibly estimate the survival function from lifetime data, such as the Kaplan--Meier estimator \cite{kaplan1958nonparametric}, or to estimate the intensity of a point process for event arrivals, such as the isotonic Hawkes process \cite{wang2016isotonic} and neural Hawkes process \cite{mei2017neural} that can be applied to the analysis of recurring events. When exploring the relationship between the covariates and time to events,
existing survival analysis methods 
often parameterize the hazard function 
with
 a weighted linear combination of covariates. One of the most popular ones is the Cox proportional hazards 
 model \cite{cox1992regression}, which is semi-parametric in that it assumes 
a non-parametric baseline hazard rate to capture the time effect. These methods are often applied to population-level studies that try to unveil the relationship between the risk factors and hazard function, such as to what degree a unit increase in a covariate is multiplicative to the hazard rate. However, the interpretability is often obtained by sacrificing model flexibility, because the proportional-hazards assumption is violated when the covariate effects are non-monotonic. For example, both very high and very low ambient temperature were related to high mortality rates in Valencia, Spain, 1991-1993 \cite{ballester1997mortality}, and a significantly increased 
mortality rate is associated with both underweight and obesity \cite{flegal2007cause}.

To accommodate nonlinear covariate effects such as non-monotonicity, existing (semi-)parametric models often expand the design matrix with transformed data, like the basis functions of smoothing splines \cite{li2005boosting,lu2008boosting} and other transformations guided by subjective knowledge. 
Instead of using hand-designed data transformations, there are several recent studies in machine learning that model complex covariate dependence with flexible functions, such as deep exponential families \cite{ranganath2016deep}, neural networks 
\cite{katzman2018deepsurv,zhu2016deep,
chapfuwa2018adversarial} and Gaussian processes \cite{fernandez2016gaussian}. With enhanced flexibilities, these recent approaches are often good at assessing individual risks, such as predicting a patient's hazard function or survival time. However, except for the Gaussian process whose results are not too difficult to interpret for low-dimensional covariates, they often have difficulty in explaining how the survival is impacted by which covariates, limiting their use in survive analysis where interpretability plays a critical role. Some approaches discretize the real-valued survival time and model the surviving on discrete time points or intervals \cite{luck2017deep,li2016multi,yu2011learning,miscouridoudeep}. They transform 
 the time-to-event modeling problem into regression, classification, or ranking ones, at the expense of losing continuity information implied by the survival time and potentially having inconsistent categories between training and testing.

In survival analysis, it is very common to have competing risks, 
in which scenario the occurrence of an event under a risk precludes events under any other risks. For example, if the event of interest is death, then all possible causes of death are competing risks to each other, since a subject that died of one cause would never die of any other cause.
Apart from modeling the time to event, in the presence of competing risks, it is also important to model the event type, or under which risk the event is likely to occur first. 
Though one can censor subjects with an occurrence of the event under a competing risk other than the risk of special interest, so that every survival model that can handle censoring is able to model competing risks, it is problematic to violate the principle of non-informative censoring \cite{kalbfleisch2011statistical, 
austin2016introduction}. The analysis of competing risks should be carefully designed and people often model two types of hazard functions, cause specific \cite{putter2007tutorial,lau2009competing} and subdistribution \cite{fine1999proportional,putter2007tutorial,lau2009competing} hazard functions. The former applies to studying etiology of diseases, while the latter is favorable when developing prediction models and risk-censoring systems 
 \cite{austin2016introduction}.

In the analysis of competing risks, there is also a trade-off between interpretability and flexibility. 
The aforementioned cause specific and subdistribution hazard functions use a Cox model with competing risk \cite{wolbers2014concordance,austin2016introduction} and a Fine-Gray subdistribution model \cite{fine1999proportional}, respectively, which are both proportional hazard models. Both models are semi-parametric, and assume
 that the hazard rate is proportional to the exponential of the inner product of the covariate and regression coefficient vectors, along with a nonparametric baseline hazard function. However, the existence of non-monotonic covariate effects can easily challenge and break the proportional-hazards assumption inherited from their corresponding single-risk model. This barrier has been surmounted by nonparametric approaches, such as random survival forests \cite{ishwaran2014random}, Gaussian processes with a single layer \cite{barrett2013gaussian} or two \cite{alaa2009deep}, and classification-based neural networks that discretize the survival time \cite{lee2018deephit}. These models are designed for competing risks, using the covariates as input and the survival times (or their monotonic transformation) or probabilities as output. Though having good model fit, the non-parametric approaches are specifically used for studies at an individual level, such as predicting the survival time, but not able to tell how the covariates affect the survival or cumulative incidence functions \cite{fine1999proportional,crowder2001classical}. Moreover, it might be questionable for Alaa and van~der Schaar \cite{alaa2009deep} to assume a normal distribution on survival times which are positive almost surely and asymmetric in general.
 
 To this end, we construct Lomax delegate racing (LDR) survival model, a gamma process based nonparametric Bayesian hierarchical model for survival analysis with competing risks. The LDR survival model utilizes the race of exponential random variables to model both the time to event and event type and subtype, and uses the summation of a potentially countably infinite number of covariate-dependent gamma random variables as the exponential distribution rate parameters. It is amenable to not only censoring data, but also missing event types or event times. Code for reproducible research is available at \href{https://github.com/zhangquan-ut/Lomax-delegate-racing-for-survival-analysis-with-competing-risks}{https://github.com/zhangquan-ut/Lomax-delegate-racing-for-survival-analysis-with-competing-risks}.

\section{Exponential racing and survival analysis}
\label{sec_preliminary}

Let $t\sim\mbox{Exp}(\lambda)$ represent an exponential distribution, with probability density function (PDF)
$
f(t\given \lambda) = \lambda e^{-\lambda t}, ~~t\in\mathbb{R}_+,
$
where $\mathbb{R}_+$ represents the nonnegative side of the real line, and $\lambda>0$ is the rate parameter such that $\E[t] = \lambda^{-1}$ and $\mbox{Var}[t] = \lambda^{-2}$. 
Shown below is a well-known property that characterizes a race among independent exponential random variables \cite{Ross:2006:IPM:1197141,caron2012bayesian}.

\begin{property}[Exponential racing
]\label{prop_expr}
If $t_j\sim\emph{\mbox{Exp}}(\lambda_j)$, where $j=1,\ldots,J$, are 
 independent to each other, 
 then $t=\min\{t_1,\ldots,t_J\}$ 
and the argument of the minimum $y=\mathop{\mathrm{argmin}}\nolimits_{j\in\{1,\ldots,J\}} t_j$ are independent, satisfying 
\beq\textstyle
t\sim \emph{ \mbox{Exp}}\left(\sum\nolimits_{j=1}^J \lambda_{j}\right), ~y\sim \emph{\mbox{Categorical}}\left({\lambda_1}\Big/{\sum\nolimits_{j=1}^J \lambda_{j}}, \cdots, {\lambda_J}\Big/{\sum\nolimits_{j=1}^J \lambda_{j}}\right).
\label{eq:Cat}
\eeq
\end{property}
Suppose there is a race among teams $j=1,\cdots,J$, whose completion times $t_j$ follow $\mbox{Exp}(\lambda_j)$, with the winner being the team with the minimum completion time. Property \ref{prop_expr} shows the winner's completion time $t$ still follows an exponential distribution and is independent of which team wins the race. In the context of survival analysis, if we consider a competing risk as a team and the latent survival time under this risk as the completion time of the team, then $t$ will be the observed time to event (or failure time) and $y$ the event type (or cause of failure). Exponential racing not only describes a natural mechanism of competing risks, but also provides an attractive modeling framework amenable to Bayesian inference, 
as 
conditioning on $\lambda_j$'s, the joint distribution of the event type $y$ 
and time to event 
$t$ becomes fully factorized as
\begin{align}
P(y,t\given \{\lambda_{j}\}_{1,J})&
=\lambda_{y} e^{-t\sum_{j=1}^J\lambda_{j}}. \label{eq:joint_EXP}
\end{align}

In survival analysis, it is rarely the case that both $y$ and $t$ are observed for all observations, and one often needs to deal with missing data and right or left censoring. We write $t\sim\mbox{Exp}_{\Psi}(\lambda)$ as a truncated exponential random variable defined by PDF $f_\Psi(t\given \lambda) = \lambda e^{-\lambda t}/\int_{\Psi}\lambda e^{-\lambda u} du,$ where $t\in\Psi$ and $\Psi$ is an open interval on $\mathbb{R}_+$ representing censoring. Concretely, $\Psi$ can be $\bm (T_{r.c.},\infty\bm )$, indicating right censoring with censoring time $T_{r.c.}$, can be $\bm (0, T_{l.c.}\bm )$, indicating left censoring with censoring time $T_{l.c.}$, or can be a more general case $\bm (T_{1}, T_{2}\bm )$, $T_2>T_1$.

 If we do not observe $y$ or $t$, or there exists censoring, we have the following two scenarios, for both of which it is necessary to introduce appropriate auxiliary variables to achieve fully factorized likelihoods: 1) If we only observe $y$ (or $t$), then we can draw $t$ (or $y$) shown in \eqref{eq:Cat} 
as an auxiliary variable, leading to the fully factorized likelihood as in \eqref{eq:joint_EXP}; 2) If we do not observe $t$ but know $t\in\Psi$ with $P(t\in\Psi\given \{\lambda_{j}\}_{1,J})=\int_{\Psi}(\sum_j\lambda_j) e^{-\sum_j\lambda_j u}du$, then we draw $t\sim \mbox{Exp}_{\Psi}(\sum\nolimits_j \lambda_j)$, resulting in the likelihood
\begin{align}\textstyle
P\left(t, t\in \Psi\given \sum\nolimits_j \lambda_{j}\right)=f_\Psi\left(t\given \sum\nolimits_j\lambda_j\right)P\left(t\in\Psi\given \sum\nolimits_j\lambda_j\right)=\left(\sum\nolimits_j\lambda_j\right) e^{-t\sum\nolimits_j\lambda_j}.
\end{align}
Together with $y$, which can be drawn by \eqref{eq:Cat} if it is missing, the likelihood $P(y,t,t\in\Psi\given \{\lambda_{j}\}_{1,J})$ becomes the same as in \eqref{eq:joint_EXP}. 
The procedure of sampling $t$ and/or $y$, generating fully factorized likelihoods under different censoring conditions, plays a crucial role as a data augmentation scheme that will be used for Bayesian inference of the proposed LDR survival model. 

In survival analysis with competing risks, one is often interested in 
modeling the dependence of the event type $y$ and failure time $t$ on 
covariates $\xv=(1,x_1,\ldots,x_V)'$. Under the exponential racing framework, one may simply let $\lambda_j = e^{\xv'\betav_j}$, where $\betav_j=(\beta_{j0},\ldots,\beta_{jV})'$ is the regression coefficient vector for the $j$th competing risk or event type.
 However, 
the hazard rate for the $j$th competing risk, expressed as $\lambda_j=e^{\xv'\betav_j}$, is restricted to be log-linear in the covariates $\xv$. This clear restriction motivates us to generalize exponential racing to Lomax racing, which can have a time-varying hazard rate for each competing risk, 
and further to Lomax delegate racing, which can use the convolution of a potentially countably infinite number of covariate-dependent gamma distributions to model each~$\lambda_j$.

\section{Lomax and Lomax delegate racings} 
\label{sec_LDR}

In this section, we generalize exponential racing to Lomax racing, which relates survival analysis with competing risks to a race of conditionally independent Lomax distributed random variables. We further generalize Lomax racing to Lomax delegate racing, which races the winners of conditionally independent Lomax racings. 
Below we first briefly review Lomax distribution.

Let $\lambda\sim \mbox{Gamma}( r,1/b)$ represent a gamma distribution 
with $\E[\lambda]=r/b$ and $\mbox{Var}[\lambda]=r/b^2$. Mixing the rate parameter of an exponential distribution with $\lambda\sim \mbox{Gamma}( r,1/b)$ leads to 
a Lomax distribution \cite{lomax1954business} $t\sim\mbox{Lomax}(r,b)$, with shape $r>0$, scale $b>0$, and PDF
$$ 
\textstyle f(t\given r,b)= \int_0^\infty \mbox{Exp}(t;\lambda) \mbox{Gamma}(\lambda; r,1/b) d\lambda = 
{rb^r}{(t+b)^{-(r+1)}},~~t\in\mathbb{R}_+.
$$ 
When $r>1$, we have $\E[t]=b/(r-1)$, and when $r>2$, we have $\mbox{Var}[t]=b^2r/[(r-1)^2(r-2)]$.
 The Lomax distribution is a heavy-tailed distribution. 
Its hazard rate and survival function can be expressed as
$ 
h(t)=
r/(t+b)
$ 
and
 $S(t) = (t+b^{-1})^{-r}$, respectively.

\subsection{Covariate-dependent Lomax racing}
We generalize covariate-dependent exponential racing by letting 
$$t_j\sim\mbox{Exp}(\lambda_j),~\lambda_j\sim{\mbox{Gamma}}(r,e^{\xv' \betav_j}).$$ Marginalizing out $\lambda_j$ leads to 
$t_j\sim{\mbox{Lomax}}(r,e^{-\xv' \betav_j}) $.
Lomax distribution 
was initially introduced to study business failures \cite{lomax1954business} and has since then been widely used to model the time to event in survival analysis \cite{myhre1982screen,howlader2002bayesian,cramer2011progressively,hemmati2017adaptive}. Previous research on this distribution \cite{al2001statistical,childs2001order,giles2013bias}, however, has been mainly focused on point estimation of parameters, without modeling covariate dependence and performing Bayesian inference. 
We define Lomax racing 
 as follows. 
\begin{definition}\label{simple_exprr}
Lomax racing models the time to event $t$ and event type $y$ given covariates $\xv$ as 
\beq
t = t_y,~y=\mathop{\mathrm{argmin}}\nolimits_{j\in\{1,\ldots,J\}} t_j, ~t_j\sim\emph{\mbox{Lomax}}(r,e^{-\xv' \betav_j}) . \label{eq:LomaxRace}
\eeq
\end{definition}
To explain the notation, suppose a patient has both diabetes ($j=1$) and cancer ($j=2$), then $t_1$ will be the patient's latent survival time under diabetes and $t_2$ under cancer. The patient's observed survival time is $\min (t_1,t_2)$. Note Lomax racing can also be considered as an exponential racing model with multiplicative random effects, since $t_j$ in \eqref{eq:LomaxRace} can also be generated as 
 $$t_j\sim\mbox{Exp}(\epsilon_j e^{\xv' \betav_j} ),~\epsilon_j\sim\mbox{Gamma}(r,1) .$$
There are two clear benefits of Lomax racing over exponential racing. The first benefit is that given $\xv$ and $\betav_j$, the hazard rate for the $j$th competing risk, expressed as $r/(t_j+e^{-\xv'\betav_j})$, is no longer a constant as $e^{\xv'\betav_j}$. The second benefit is that closed-form Gibbs sampling update equations can be derived, as will be described in detail in Section \ref{sec:inference} and the Appendix. 

For competing risk $j$, we can also express $t_j\sim\mbox{Exp}(\epsilon_j e^{\xv' \betav_j} ),~\epsilon_j\sim\mbox{Gamma}(r,1) $ as
\beq
\ln(t_j)= -\xv'\betav_j+ \varepsilon_j,~\varepsilon_j =\ln (\varepsilon_{j1}/ \varepsilon_{j2}),~\varepsilon_{j1} \sim \mbox{Exp}(1), ~\varepsilon_{j2} \sim \mbox{Gamma}(r,1). \notag
\eeq
Thus Lomax racing regression uses an accelerated failure time model \cite{kalbfleisch2011statistical} for each of its competing risks. More specifically, with
$S_0(t_j) = (t_j+1)^{-r}$ and $h_0(t_j) = \frac{r}{t_j+1}$, we have
\beq
S_j(t_j) = (e^{\xv'\betav_j}t_j+1)^{-r} = S_0(e^{\xv'\betav_j}t_j),~~h_j(t_j) = {r}{(t_j+e^{-\xv'\betav_j})^{-1}} 
= e^{\xv'\betav_j} h_0(e^{\xv'\betav_j}t_j ), \label{eq:lr_risk_func}
\eeq
and hence $e^{-\xv'\betav_j}$ can be considered as the accelerating factor for competing risk $j$. 
Considering all $J$ risks, we can express survival function $S(t)$ and hazard function $h(t)$ as
\beq
\small S(t) =\prod_{j=1}^J S_j(t)=\prod_{j=1}^J (e^{\xv'\betav_j}t+1)^{-r} = \prod_{j=1}^J S_0(e^{\xv'\betav_j}t),~~~~h(t) = \frac{-d S(t)/dt}{S(t)} = \sum_{j=1}^J\frac{r}{t+e^{-\xv'\betav_{j}}}.\label{eq:lr_func}
\eeq

The nosology of competing risks is often subjected to human knowledge, diagnostic techniques, and patient population. Diseases with the same phenotype, categorized into one competing risk, might have distinct etiology and different impacts on survival, and thus require different therapies. For example, for a patient with both diabetes and cancer, it can be unknown whether the patient has Type 1 or Type 2 diabetes, where
Type 1 
is ascribed to insufficient production of insulin from pancreas whereas Type 2 arises from the cells' failure in responding properly to insulin \cite{varma2014prevalence}. In this regard, it is often necessary for a model to identify sub-risks within a pre-specified competing risk, which may not only improve the fit of survival time, but also help diagnose new disease subtypes. We develop Lomax delegate racing, assuming that a risk consists of several sub-risks, under each of which the latent failure time is accelerated by the exponential of a weighted linear combination of covariates.

\subsection{Lomax delegate racing} 
Based on the idea of Lomax racing that an individual's observed failure time is the minimum of latent failure times under competing risks, we further propose \textit{Lomax delegate racing} (LDR), assuming a latent failure time under a competing risk is the minimum of the failure times under a number of sub-risks appertaining to this competing risk. In particular, let us first denote $G_j\sim\Gamma\mbox{P}(G_{0j},1/c_{0j})$ as a gamma process defined on the product space $\mathbb{R}^+\times\Omega$, where $\mathbb{R}^+=\{x:x>0\}$, $G_{0j}$ is a finite and continuous base measure over a complete separable metric space $\Omega$, and $1/c_{0j}$ is a positive scale parameter, such that $G_j(A)\sim\mbox{Gamma}(G_{0j}(A), 1/c_{0j})$ for each Borel set $A\subset \Omega$. A draw from the gamma process consists of countably infinite non-negatively weighted atoms, expressed as $G_j=\sum_{k=1}^\infty r_{jk} \delta_{\betav_{jk}}$. Now we formally define LDR survival model as follows.

\begin{definition}[Lomax delegate racing]\label{bnpexprr}
Given a random draw of a gamma process $G_j\sim\Gamma\emph{\mbox{P}}(G_{0j},1/c_{0j})$, expressed as $G_j=\sum_{k=1}^\infty r_{jk} \delta_{\betav_{jk}}$, for each $j\in\{1,\ldots,J\}$, Lomax delegate racing models
the time to event $t$ and event type $y$ given covariates $\xv$ as
\beq
t=t_y,~y=\mathop{\mathrm{argmin}}\limits_{j\in\{1,\ldots,J\}} t_{j}, ~ t_j = t_{j\kappa_j},~ 
\kappa_j=\mathop{\mathrm{argmin}}\limits_{k\in\{1,\ldots,\infty\}} t_{jk}, ~t_{jk}\sim\emph{\mbox{Lomax}}(r_{jk},e^{-\xv' \betav_{jk}}). \label{eq:argmin_argmin}
\eeq
\end{definition}\vspace{-3mm}
In contrast to specifying a fixed number of competing risks $J$, the gamma process not only admits a race among a potentially infinite number of sub-risks, but also parsimoniously shrinks toward zero the weights of negligible sub-risks \cite{zhou2016augmentable, SoftplusReg_2016}, so that the non-monotonic covariate effects on the failure time under a competing risk can be interpreted as the \textit{minimum}, which is a nonlinear operation, of failure times under sub-risks whose accelerating factor is log-linear in $\xv$.
As shown in the following Corollary, 
LDR can also be considered as a generalization of exponential racing, where the exponential rate parameter of each competing risk $j$ is a weighted summation of a countably infinite number of gamma random variables with covariate-dependent weights. 
\begin{cor}\label{cor:Exp_sum_gamma}
Lomax delegate racing survival model can also be expressed as
\begin{align}
t = t_y,~y=\mathop{\mathrm{argmin}}\nolimits_{j\in\{1,\ldots,J\}} t_{j},~ t_j\sim\emph{\mbox{Exp}}\left(\sum\nolimits_{k=1}^\infty e^{\xv'\betav_{jk}}\tilde\lambda_{jk}\right),~ \tilde\lambda_{jk} \sim\emph{\mbox{Gamma}}(r_{jk}, 1).
\label{eq:argmin_Exp}
\end{align}
\end{cor}

We provide in the Appendix the marginal distribution of $t$ in LDR for situations where predicting the failure time is of interest. The survival and hazard functions of LDR, which generalize those of Lomax racing in \eqref{eq:lr_func}, can be expressed as
\small
\begin{align}
&S(t) = \prod_{j=1}^J\prod_{k=1}^\infty P(T_{jk}>t_j) = \prod_{j=1}^J\prod_{k=1}^\infty (e^{\xv'\betav_{jk}}t_j+1)^{-r_{jk}} ,~~~~h(t) = \sum_{j=1}^J \sum_{k=1}^\infty \frac{r_{jk} }{t_j+e^{-\xv'\betav_{jk}}}.
\end{align}\normalsize

LDR survival model can be considered as a two-phase racing, where in the first phase, for each of the $J$ pre-specified competing risk there is a race among countably infinite sub-risks, and in the second phase, $J$ risk-specific failure times race with each other to eventually determine both the observed failure time $t$ and event type $y$. 
Moreover, Corollary \ref{cor:Exp_sum_gamma}, representing LDR as a single-phase exponential racing, more explicitly explains non-monotonic covariate effects on $t_j$ by writing the exponential rate parameter of $t_j$ as the aggregation of $\{e^{\xv'\betav_{jk}}\}_{k=1}^\infty$ weighted by gamma random variables with the shape parameters as the atom weights of the gamma process $G_j$.

\section{Bayesian inference}\label{sec:inference} 

LDR utilizes a gamma process \cite{ferguson73} to support countably infinite regression-coefficient vectors for each pre-specified risk. 
The gamma process $G_j\sim\Gamma\mbox{P}(G_{0j},1/c_{0j})$ has an inherent shrinkage mechanism in that the number of atoms whose weights are larger than a positive constant $\epsilon$ is finite almost surely and follows a Poisson distribution with mean $\int_{\epsilon}^\infty r^{-1}e^{-c_{0j}r}dr$. For the convenience of implementation, as in Zhou et al. \cite{NBP2012}, we truncate the total number of atoms of a gamma process to be $K$ by choosing a finite and discrete base measure as $G_{0j}=\sum_{k=1}^K \frac{\gamma_{0j}}{ K}\delta_{\betav_{jk}}$. Let us denote $\xv_i$ and $y_i$ as the covariates and the event type, respectively, for individual $i\in\{1,\ldots,n\}$. 
We express the full hierarchical model of the (truncated) gamma process LDR survival model as
\begin{align}
&y_i=\mathop{\mathrm{argmin}}\limits_{j\in\{1,\ldots,J\}} t_{ij},~~t_i=\min_j t_{ij}, ~~ 
t_{ij}= t_{ij\kappa_{ij}},~~\kappa_{ij}= \mathop{\mathrm{argmin}}\limits_{k\in\{1,\ldots,K\}} t_{ijk},~~t_{ijk} \sim \mbox{Exp}(\lambda_{ijk}),
 \nonumber\\ 
&\lambda_{ijk} \sim \mbox{Gamma}(r_{jk}, e^{\xv_{i}'\betav_{jk}}),~~r_{jk}\sim \mbox{Gamma}(\gamma_{0j}/K, 1/c_{0j}),~
\mbox{ } \betav_{jk}\sim \prod\nolimits_{v=0}^{V} \mathcal{N}(0,\alpha_{vjk}^{-1}), 
\label{hier_classifier}
\end{align}
where we further let $\alpha_{vjk}\sim \mbox{Gamma}(a_0,1/b_0)$.
The joint probability given $\{\lambda_{ijk}\}_{jk}$ is
\begin{align}
P(t_i, \kappa_{iy_i}, y_i\given \{\lambda_{ijk}\}_{jk})&=P(t_i\given \{\lambda_{ijk}\}_{jk})P( \kappa_{iy_i}, y_i\given \{\lambda_{ijk}\}_{jk})= \lambda_{i y_i \kappa_{i y_i}} e^{-t_i\sum_{j=1}^S\sum_{k=1}^K\lambda_{ijk}}, \notag
\end{align}
which is amenable to posterior simulation for $\lambda_{ijk}$.
Let us denote $\mbox{NB}(x;r,p)=\frac{\Gamma(x+r)}{x!\Gamma(r)}p^x(1-p)^r$ as the likelihood for negative binomial distribution and $\sigma(x)=1/(1+e^{-x})$ as the sigmoid function.
Further marginalize out $ \lambda_{ijk} \sim{\mbox{Gamma}}(r_{jk}, e^{\xv_i'\betav_{jk}})$ leads to a fully factorized joint likelihood as
\begin{align}
&\!\!\small P(t_i, \kappa_{iy_i}, y_i\given \xv_i,\{\betav_{jk}\}_{jk})
=t_i^{-1} \prod\nolimits_{j}\prod\nolimits_k \mbox{NB}\left(\mathbf{1}(\kappa_{iy_i} = k,y_i=j); r_{jk}, \sigma(\xv_i'\betav_{jk}+\ln t_i )
\right), \label{eq:jointNB}
\end{align}
which is amenable to 
posterior simulation 
using the data augmentation based inference technique for negative binomial regression \cite{LGNB_ICML2012,polson2013bayesian}.
The augmentation schemes of $t_i$ and/or $y_i$ discussed in Section~\ref{sec_preliminary} are used to achieve \eqref{eq:jointNB} in the presence of censoring or as a remedy for missing data.
We describe in detail both Gibbs sampling and maximum a posteriori (MAP) inference in the Appendix.

\section{Experimental results}\label{sec_experiments}

\begin{table}[t!]
\centering
\caption{Synthetic data generating process.}
\label{tab:synthetic}
\begin{tabular}{c|c}
\toprule 
Synthetic data 1 & Synthetic data 2 \\\midrule
 $t_i=\min(t_{i1},t_{i2}, 3.5)$, &$t_i=\min(t_{i1},t_{i2}, 6.5)$,\\
$t_{i1}\sim\mbox{Exp}(e^{\xv_i' \betav_1})$,~$t_{i2}\sim\mbox{Exp}(e^{\xv_i' \betav_2})$ & $t_{i1}\sim\mbox{Exp}(1/\cosh(\xv_i' \betav_1))$,~$t_{i2}\sim\mbox{Exp}(1/|\sinh(\xv_i' \betav_2)|)$\\
%
 \bottomrule
\end{tabular}\vspace{-0.5mm}
\end{table}

\begin{figure}[th!]\vspace{0mm}
 \centering
 \begin{subfigure}[t]{0.25\textwidth}
 \centering
\includegraphics[width=1\linewidth]{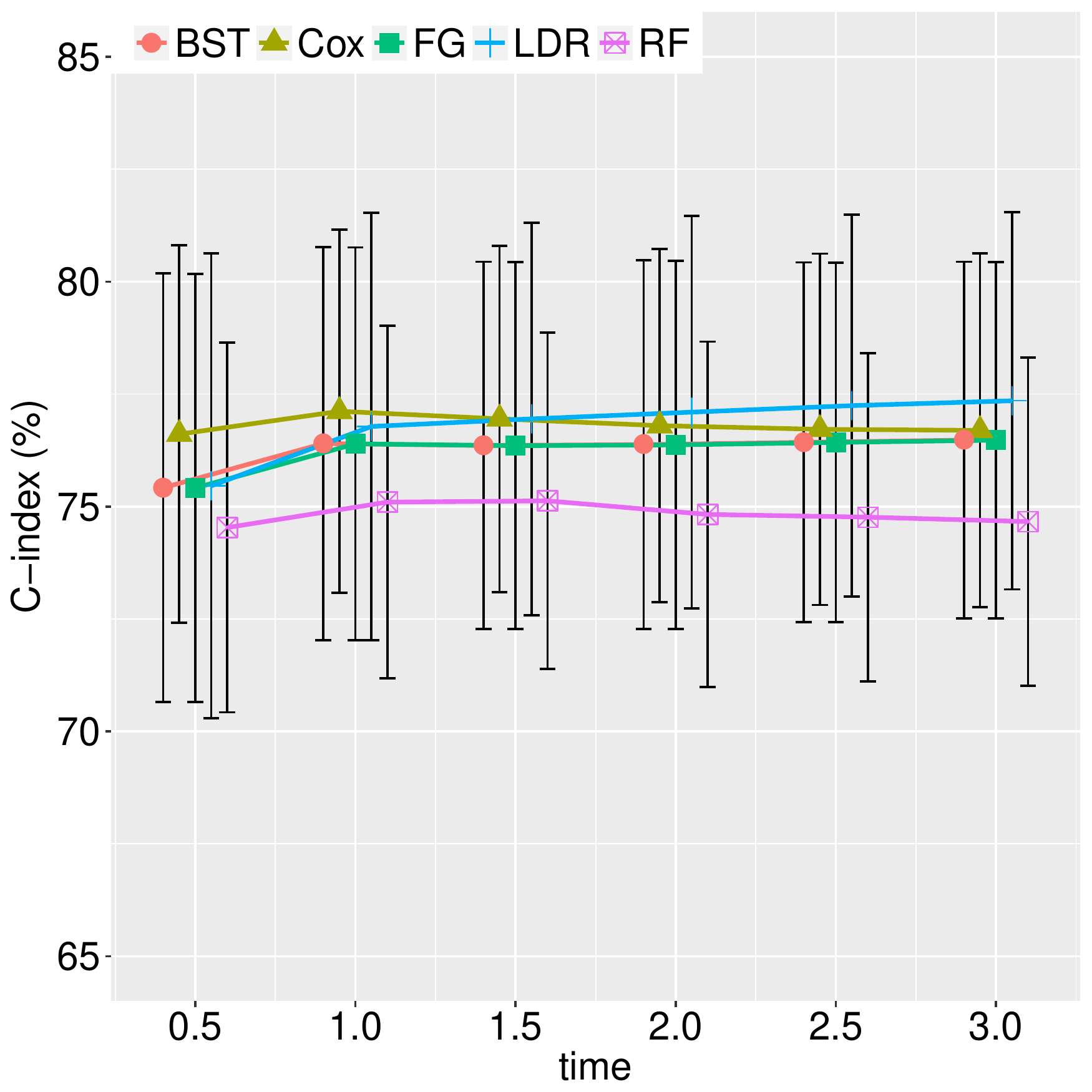}\vspace{-2.5mm}\label{linear_risk1}
 \caption{C-index of risk 1 for \\\mbox{ } \mbox{ } \, synthetic data 1.}\vspace{-1mm}
 \end{subfigure}%
 \begin{subfigure}[t]{0.25\textwidth}
 \centering
\includegraphics[width=1\linewidth]{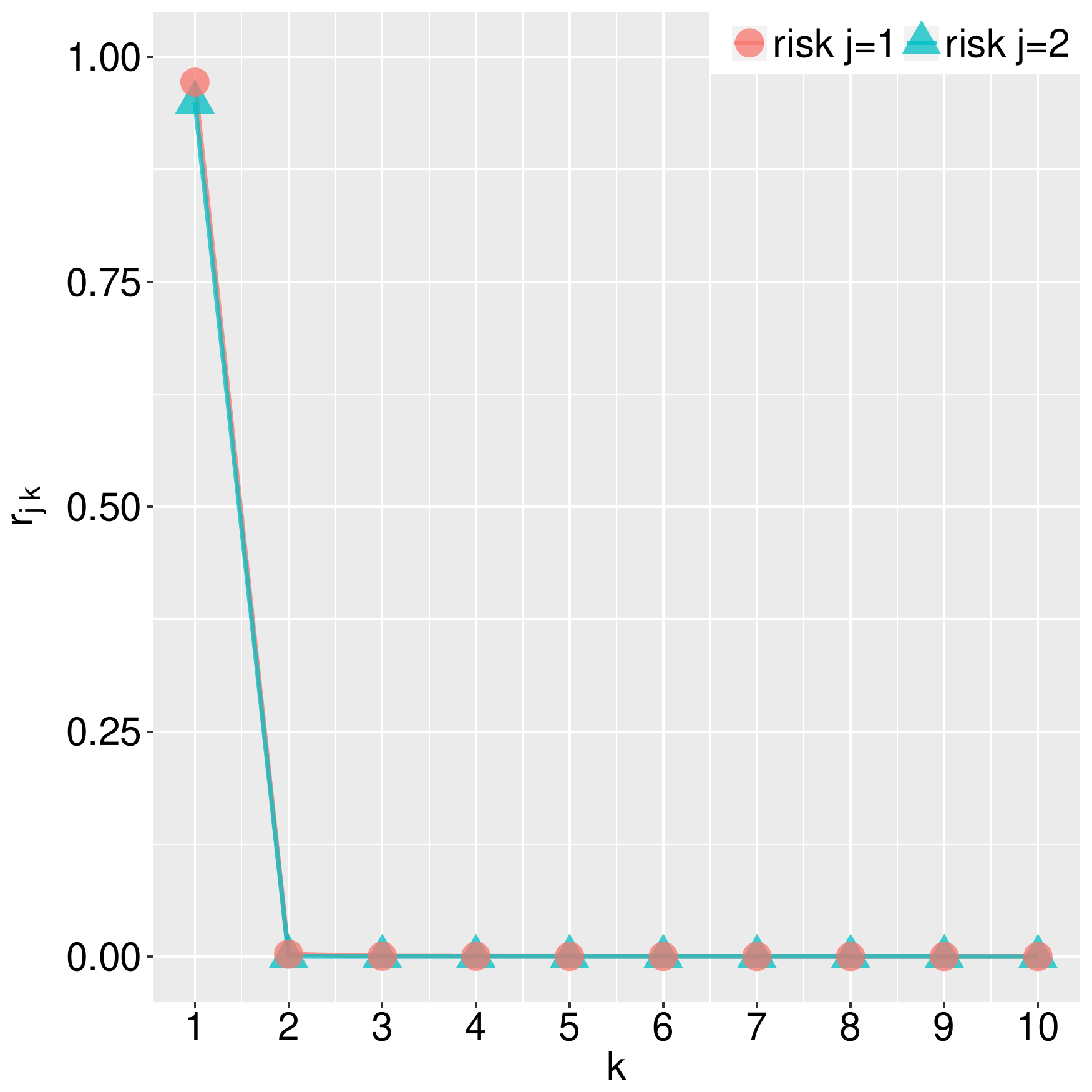}\vspace{-2.5mm}\label{linear_r}
 \caption{$r_{jk}$ by descending or-\\
 \mbox{ } \, der for synthetic data 1.
 }\vspace{-1mm}
 \end{subfigure}%
\begin{subfigure}[t]{0.25\textwidth}
 \centering
\includegraphics[width=1\linewidth]{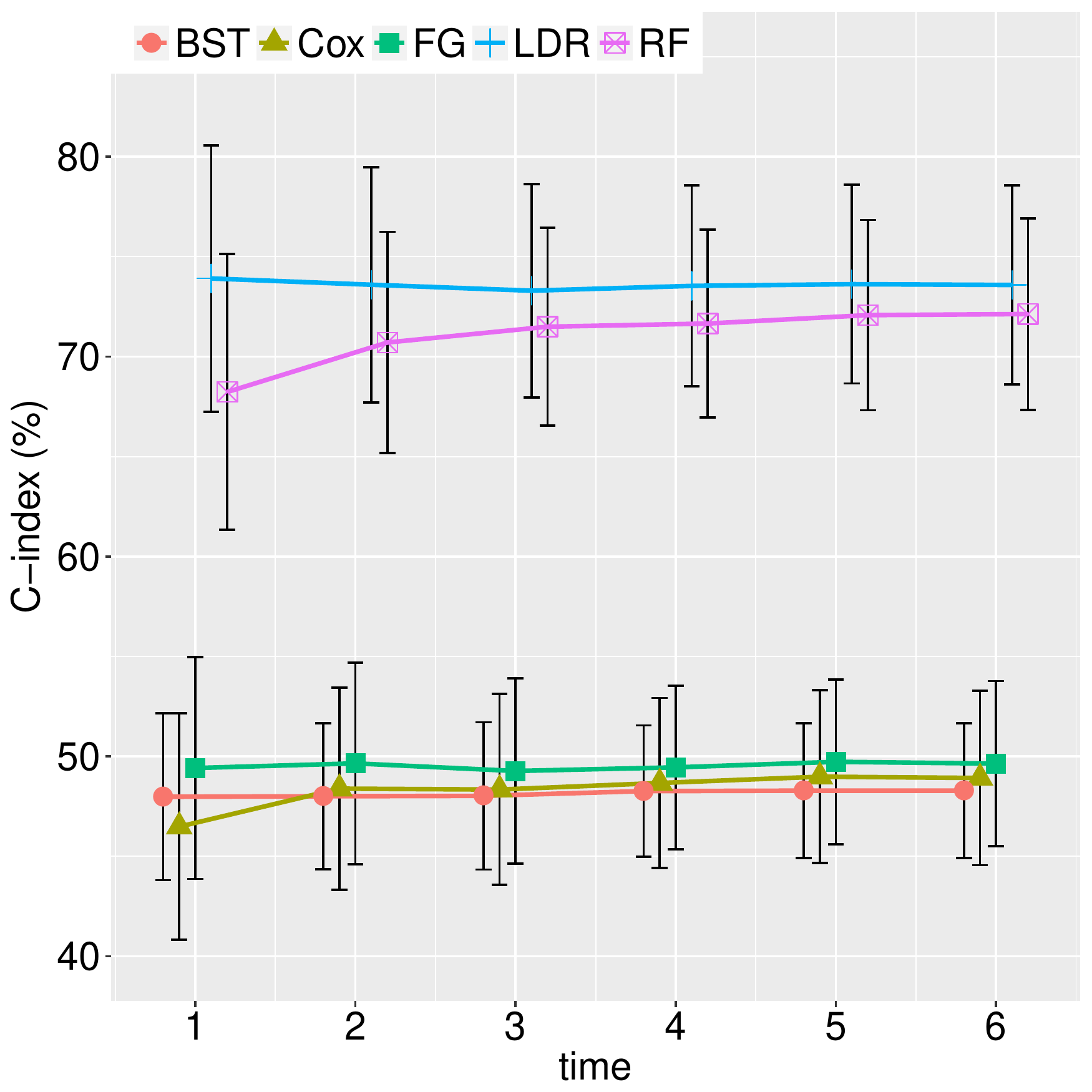}\vspace{-2.5mm}\label{cosh_risk1}
 \caption{C-index of risk 1 for \\\mbox{ } \mbox{ } \, synthetic data 2.}\vspace{-1mm}
 \end{subfigure}%
 \begin{subfigure}[t]{0.25\textwidth}
 \centering
\includegraphics[width=1\linewidth]{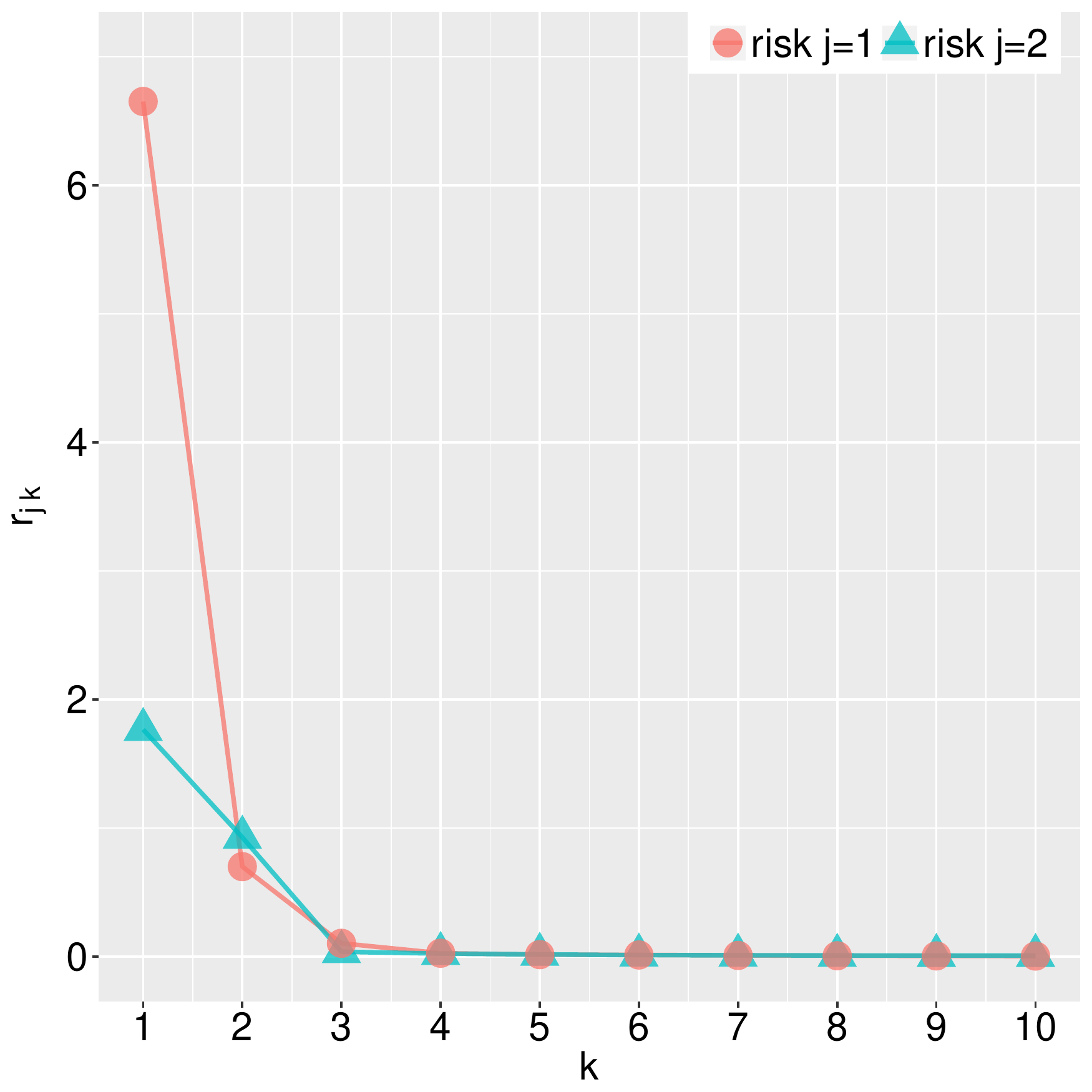}\vspace{-2.5mm}\label{cosh_r}
 \caption{$r_{jk}$ by descending order \\\mbox{ } \mbox{ } \, for synthetic data 2.
 }\vspace{-1mm}
 \end{subfigure}
\caption{Cause-specific C-indices and shrinkage of $r_{jk}$ by LDR for synthetic data 1 and 2.
 }\label{linear_toy}\vspace{-1mm}
\end{figure}

In this section, we validate the proposed LDR model by a variety of experiments using both synthetic and real data. Some data description, implementation of benchmark approaches, and experiment settings are deferred to the Appendix for brevity. 
In all experiments we exclude from the testing data the observations that have unknown failure times or event types. We compare the proposed LDR survival model, 
 cause-specific Cox proportional hazards model (Cox)\cite{wolbers2014concordance,austin2016introduction}, Fine-Gray proportional subdistribution hazards model (FG) \cite{fine1999proportional} and its boosting algorithm (BST) which is more stable for high-dimensional covariates \cite{binder2009boosting}, and random survival forests (RF) \cite{ishwaran2014random}, which are all designed for survival analysis with competing risks. We show that LDR performs uniformly well regardless of whether the covariate effects are monotonic or not. Moreover, LDR is able to infer the missing cause of death and/or survival time of an observation, both of which in general cannot be handled by these benchmark methods. { The model fits of LDR by Bayesian inference via Gibbs sampling and MAP inference via stochastic gradient descent (SGD) are comparable. We will report the results of Gibbs sampling, as it provides an explicit criterion to prune unneeded model capacity (Steps 1 and 8 of Appendix B), avoiding the need of model selection and parameter tuning. For large scale data, performing MAP inference via SGD is recommended if Gibbs sampling takes too long to run a sufficiently large number of iterations.}
We quantify model performance by cause-specific concordance index (C-index) \cite{wolbers2014concordance}, 
 where the C-index of risk $j$ at time $\tau$ in this paper is computed as
\begin{align*}
\mathcal{C}_{j}(\tau)=P\left(
Score_j(\xv_i,\tau) > Score_j(\xv_{i'},\tau)
\given y_i=j \mbox{ and } [t_i<t_{i'} \mbox{ or } y_{i'}\neq j] \right),
\end{align*}
where $i\neq i'$ and 
 $Score_j(\xv_i, \tau)$ is a prognostic score at time $\tau$ depending on $\xv_i$ such that its higher value reflects a higher risk of cause $j$. 
Intuitively, for cause $j$, if patient $i$ died of this cause ($i.e.$, $y_i=j$), and patient $i'$ either died of another cause ($i.e.$, $y_{i'}\neq j$) or died of this cause but lived longer than patient $i$ ($i.e.$, $t_i<t_{i'}$), then it is likely that $Score_j(\xv_i, \tau)$ for patient~$i$ is higher than $Score_j(\xv_{i'}, \tau)$ for patient $i'$, and the ranking of risks for this pair of patients is concordant. C-index measures such concordance, and a higher value indicates better model performance.
Wolbers et al. \cite{wolbers2014concordance} write C-index as a weighted average of time-dependent AUC that is related to sensitivity, specificity, and ROC curves for competing risks \cite{saha2010time}. So a C-index around $0.5$ implies a model failure. 
A good choice of the prognostic score is the cumulative incidence function, $i.e,$ $ Score_j(\xv_i,\tau)=\mbox{CIF}_j(i,\tau)=P(t_i\leq \tau, y_i=j)$ \cite{fine1999proportional,kalbfleisch2011statistical,crowder2001classical}. 
Distinct from a survival function that measures the probability of surviving beyond some time, CIF estimates the probability that an event occurs by a specific time in the presence of competing risks. 
For LDR given $\{r_{jk}\}$ and $\{\betav_{jk}\}$,
\begin{align*}\textstyle
\mbox{CIF}_j(i,\tau)=P(t_i\leq \tau, y_i=j)=\E\left[ \frac{\sum_{k}\lambda_{ijk}}{\sum_{j',k}\lambda_{ij'k}}\left(1-e^{-\tau\sum_{j',k}\lambda_{ij'k} } \right)\right],
\end{align*}
where the expectation is taken over $\{\lambda_{jk}\}_{j,k}$, where $\lambda_{ijk}\sim \mbox{Gamma}(r_{jk},e^{\xv_i'\betav_{jk}})$. The expectation can be evaluated by Monte-Carlo estimation 
 if we have a point estimate or a collection of post-burn-in MCMC samples of $r_{jk}$ and $\betav_{jk}$.

\subsection{Synthetic data analysis}\label{sec:synthetic}
We first simulate two datasets following Table \ref{tab:synthetic}, where $\bm x_i\sim \mbox{N}(\bm0, \Imat_3)$, to illustrate the unique nonlinear modeling capability of LDR. In Table \ref{tab:synthetic} $t_{ij}$ denotes the latent survival time under risk $j$, $j=1,2$ and $t_i$ is the observed time to event. The event type $y_i=\arg\min_j t_{ij}$ if $t_i < T_{r.c.}$, with $y_i=0$ indicating right censoring if $t_i=T_{r.c.}$, where the censoring time $T_{r.c.}=3.5$ for data 1 and $6.5$ for data 2. We simulate 1,000 random observations, and use 800 for training and the remaining 200 for testing. We randomly take 20 training/testing partitions, on each of which we evaluate the testing cause-specific C-index at time $0.5,1, 1.5, \cdots, 3$ for data 1 and at time $1,2,\cdots, 6$ for data 2. The sample mean $\pm$ standard deviation of the estimated cause-specific C-indices of risks 1, and the estimated $r_{jk}$'s of both risks by LDR (from one random training/testing partition but without loss of generality) for data 1 are displayed in panels (a) and (b) of Figure \ref{linear_toy}, respectively. Analogous plots for data 2 
are shown in panels (c) and (d). The testing C-indices of risk 2 are analogous to those of risk 1 for both datasets, thus shown in Figure \ref{toyrisk2} in the Appendix for brevity.

For data 1 where the survival times under both risks depend on the covariates monotonically, LDR has comparable performance with Cox, FG, and BST, and all these four models slightly outperform RF in terms of the mean values of C-indices. The underperformance of RF in the case of monotonic covariate effects has also been observed in its original paper \cite{ishwaran2014random}. For data 2 where the survival time and covariates are not monotonically related, 
 LDR and RF at any time evaluated significantly outperform the other three approaches, all of which fail on this dataset as their C-indices are around $0.5$ for both risks. Panels (b) and (d) of Figure \ref{linear_toy} show $r_{jk}$ 
 inferred on data 1 and 2, respectively. More specifically, both risks consist of only one sub-risk for data 1. By contrast, two sub-risks of the two respective risks can approximate the complex data generating process of data 2.

\subsection{Real data analysis}

\begin{figure}[t!]
 \centering
 \begin{subfigure}[t]{0.26\textwidth}
 \centering
\includegraphics[width=1\linewidth]{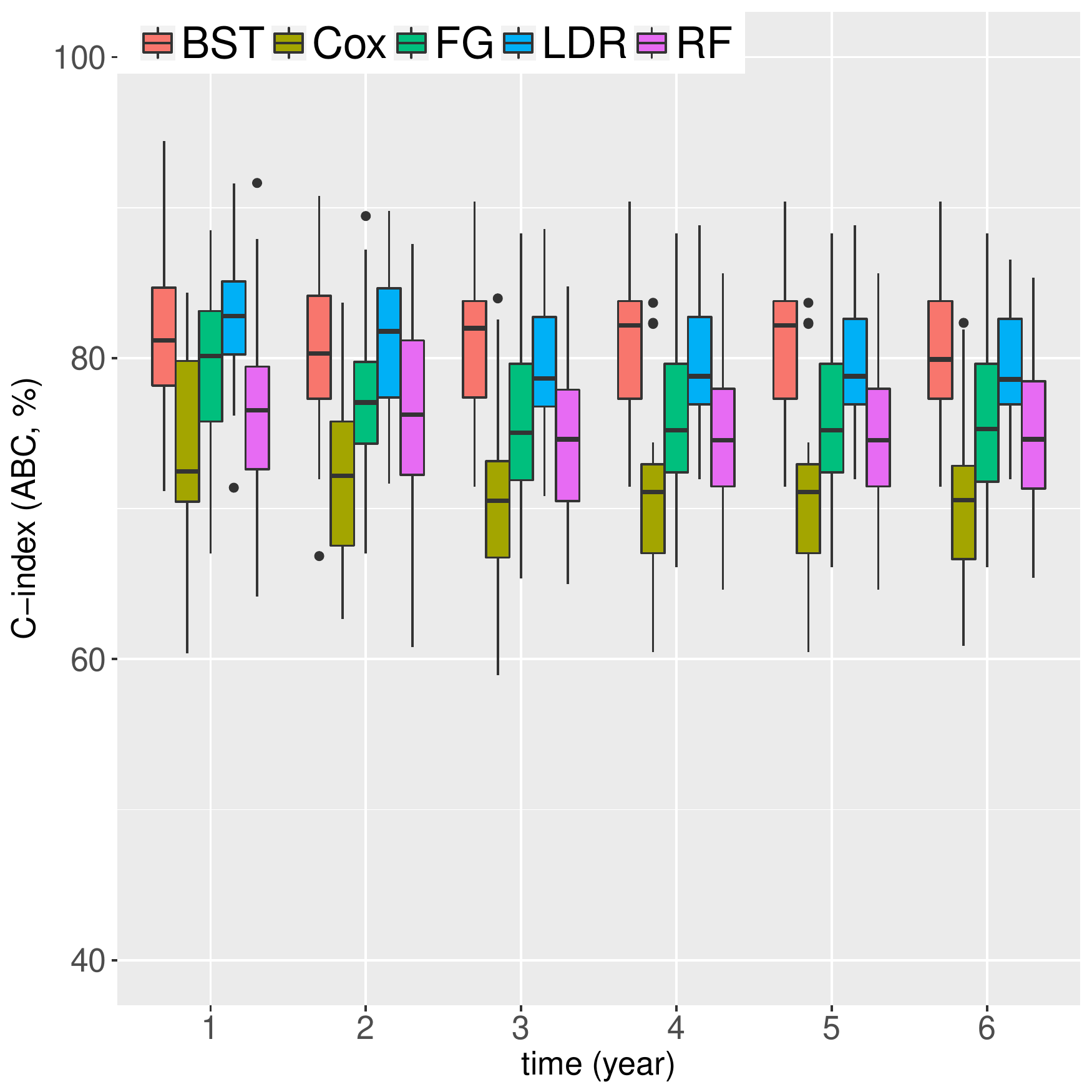}\vspace{-2mm}\label{cosh_risk1}
 \caption{C-index of ABC.}\vspace{-1.5mm}
 \end{subfigure}%
 ~ 
 \begin{subfigure}[t]{0.26\textwidth}
 \centering
\includegraphics[width=1\linewidth]{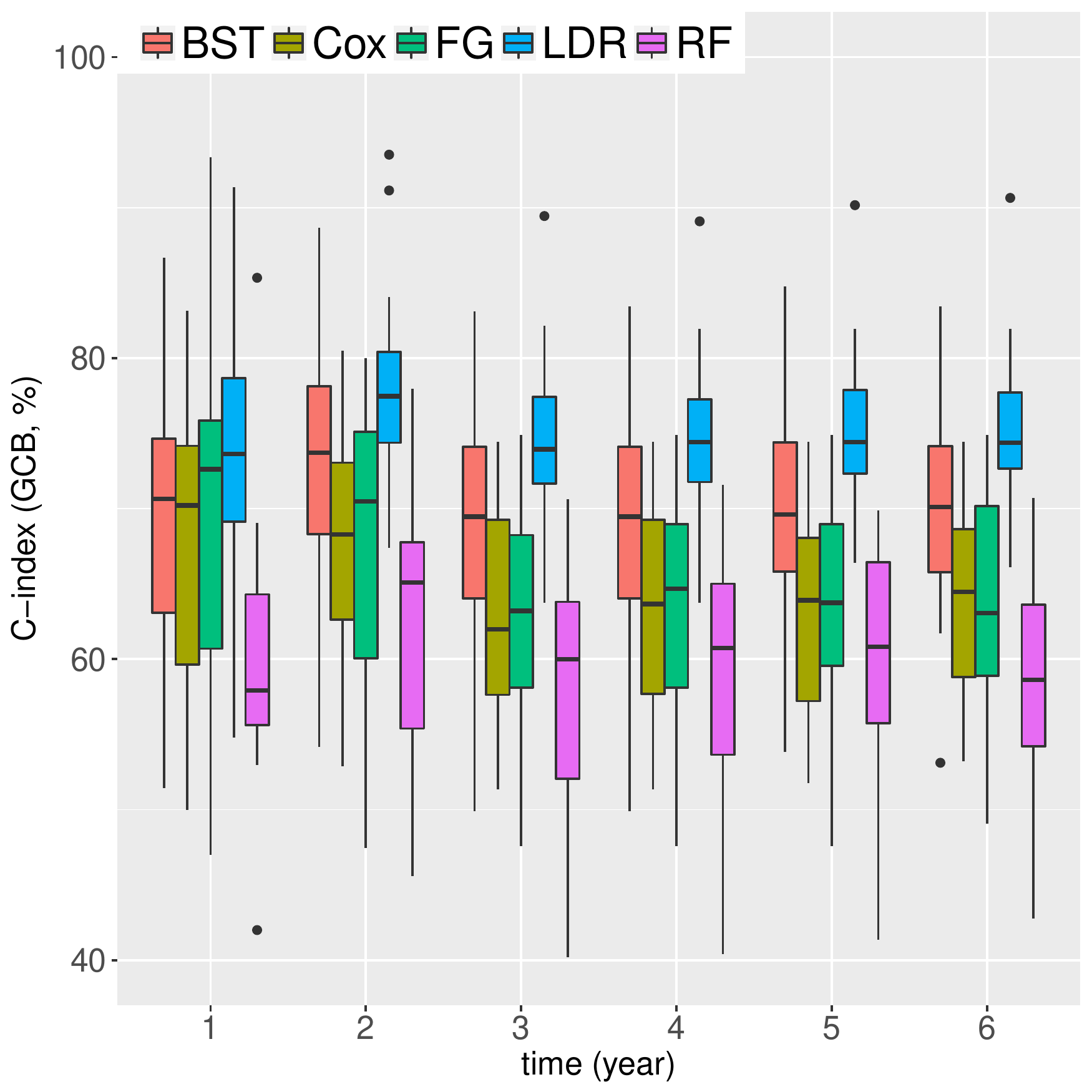}\vspace{-2mm}\label{cosh_risk2}
 \caption{C-index of GCB.}\vspace{-1.5mm}
 \end{subfigure}%
 ~ 
 \begin{subfigure}[t]{0.26\textwidth}
 \centering
\includegraphics[width=1\linewidth]{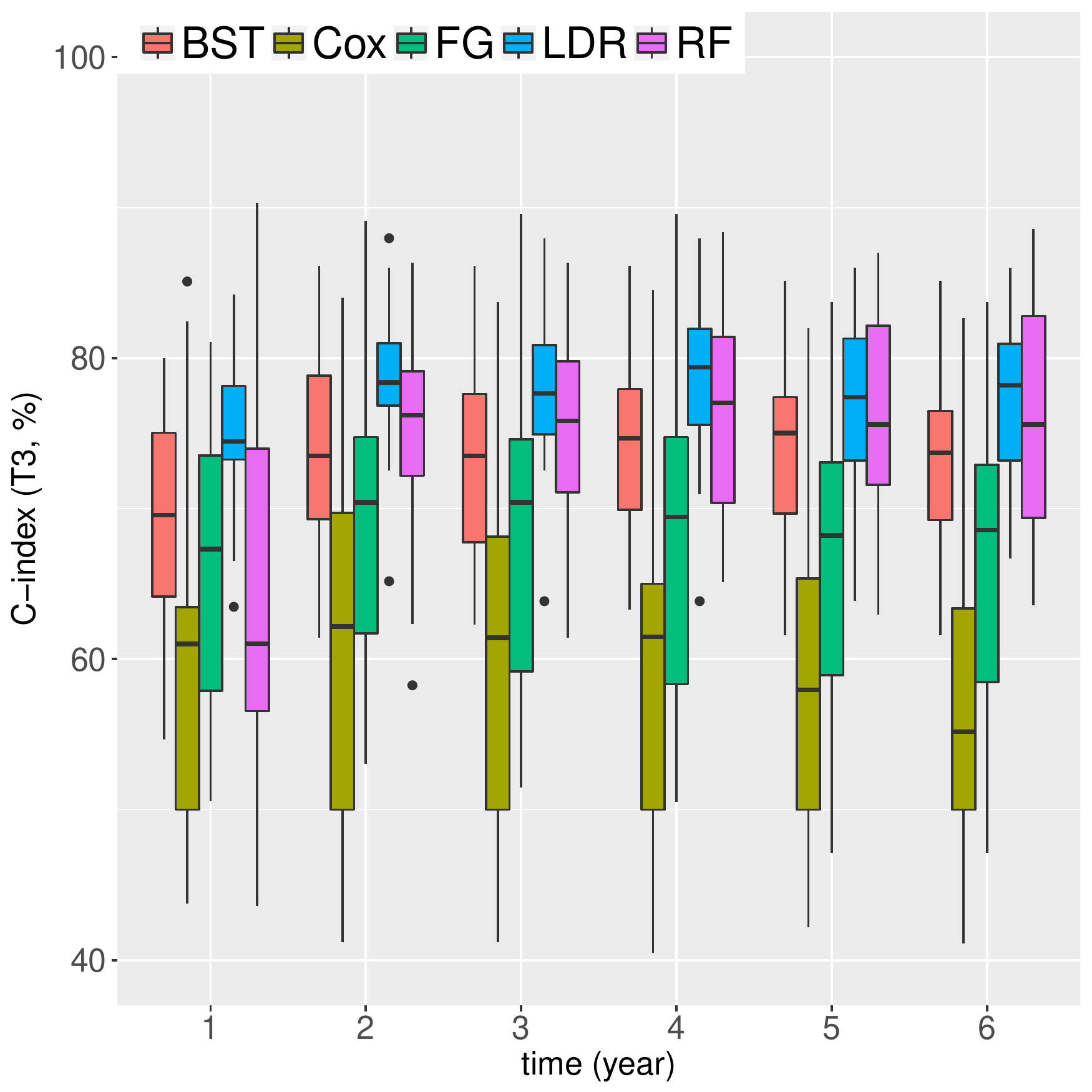}\vspace{-2mm}\label{cosh_r}
 \caption{C-index of T3.}\vspace{-1.5mm}
 \end{subfigure}%
\caption{Cause-specific C-indices for DLBCL data.
 }\label{lymphoma_cindex}\vspace{-2.mm}
\end{figure}

We analyze a microarray gene-expression profile \cite{rosenwald2002use} to assess our model performance on real data. The dataset contains a total of 240 patients with diffuse large B-cell lymphoma (DLBCL). Multiple unsuccessful treatments to increase the survival rate suggest that there exist several subtypes of DLBCL that differ in responsiveness to chemotherapy. In the DLBCL dataset, Rosenwald et al. \cite{rosenwald2002use} identify three gene-expression subgroups, including activated B-cell-like (ABC), germinal-center B-cell-like (GCB), and type 3 (T3) DLBCL, which may be related to three different diseases as a result of distinct mechanisms of malignant transformation. They also suspect that T3 may be associated with more than one such mechanism. In our analysis, we treat the three subgroups and their potential malignant transformation mechanisms as competing risks from which the patients suffer. As the total number of patients is small which is often the case in survival data, we consider 434 genes that have no missing values across all the patients. Seven of the 434 genes have been reported to be related to clinical phenotypes and four of the seven to have non-monotonic effects on the risk of death \cite{li2005boosting}. Since some gene expressions may be highly correlated, we follow the same selection procedure of Li and Luan \cite{li2005boosting} to include as covariates the seven genes, together with another 33 genes having the highest Cox partial score statistic, so that both Cox proportional model and FG subdistribution model for competing risks do not collapse for computational singularity. We use 200 observations  for training and the remaining 40 for testing. We take 20 random training/testing partitions and report in Figure \ref{lymphoma_cindex} boxplots of the testing C-indices evaluated at year $1,2,\cdots,6$, by the same five approaches used in the analysis of synthetic datasets.

The boxplots of BST and LDR are roughly comparable for ABC, but the median of LDR is slightly higher than those of BST until year 2, and hereafter slightly lower. For GCB and T3, LDR results in higher median C-indices than all the other benchmarks do at any time evaluated, indicating LDR provides a big improvement in predicting lymphoma CIFs. Interestingly noted is that RF has low performance in both ABC and GCB, but outperforms Cox, FG, and BST and is comparable to LDR in T3. This implies that the gene expressions may have monotonic effects on survival under ABC or GCB, but it is not the case for T3, which can be validated by the fact that LDR learns one sub-risk for ABC and GCB, respectively, and two sub-risks for T3. 
To better show the improvements of LDR over existing approaches, 
we calculate the difference of C-indices between LDR and each of the other four benchmarks within each training-testing partition, and report the sample mean and standard deviation across partitions in Table \ref{tab:lymphoma} in the Appendix. On average, the improvements of LDR over Cox, FG, and BST are bigger for T3 than those for ABC or GCB, whereas LDR outperforms RF by a larger margin for ABC and GCB than for T3. This shows another advantage of LDR that it fits consistently well regardless of whether the covariate effects are monotonic or not. 

We further analyze a publicly accessible dataset from the \textit{Surveillance, Epidemiology, and End Results} (SEER) Program of National Cancer Institute \cite{seer}. The SEER dataset we use contains two risks: one is breast cancer and the other is ``other causes,'' which we denote as BC and OC, respectively. It also contains some incomplete observations, each of which with an unknown cause of death but observed uncensored time to death, that can be handled by LDR. The individual covariates include the patients' personal information, such as age, gender, race, and diagnostic and therapy information. More details are deferred to the Appendix.

\begin{figure}[!t]\vspace{-2mm}
 \centering
 \begin{subfigure}[t]{0.24\textwidth}
 \centering
\includegraphics[width=1\linewidth]{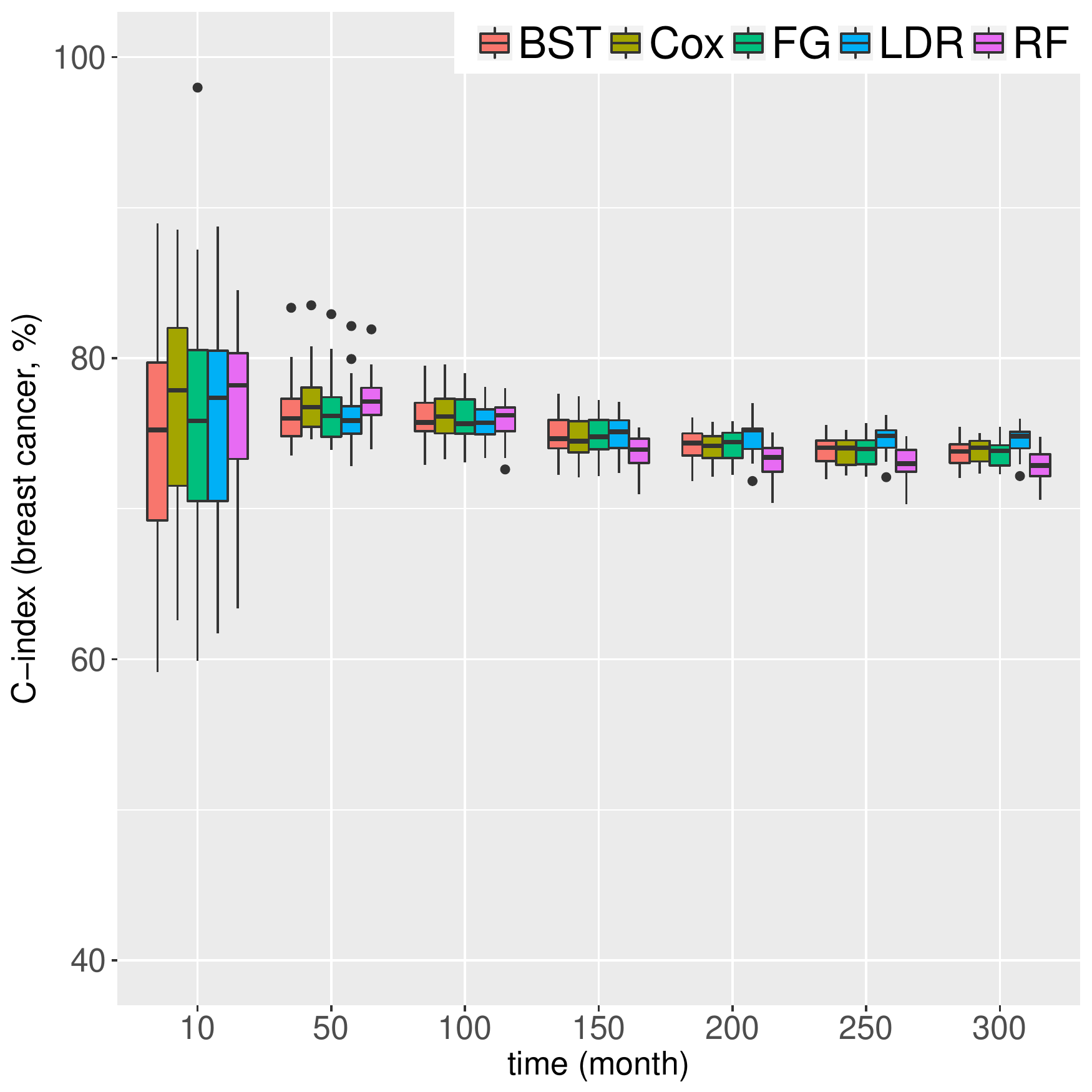}\vspace{-3mm}\label{nomissing_1}
 \caption{BC.}\vspace{-2.5mm}
 \end{subfigure}%
 ~ 
 \begin{subfigure}[t]{0.24\textwidth}
 \centering
\includegraphics[width=1\linewidth]{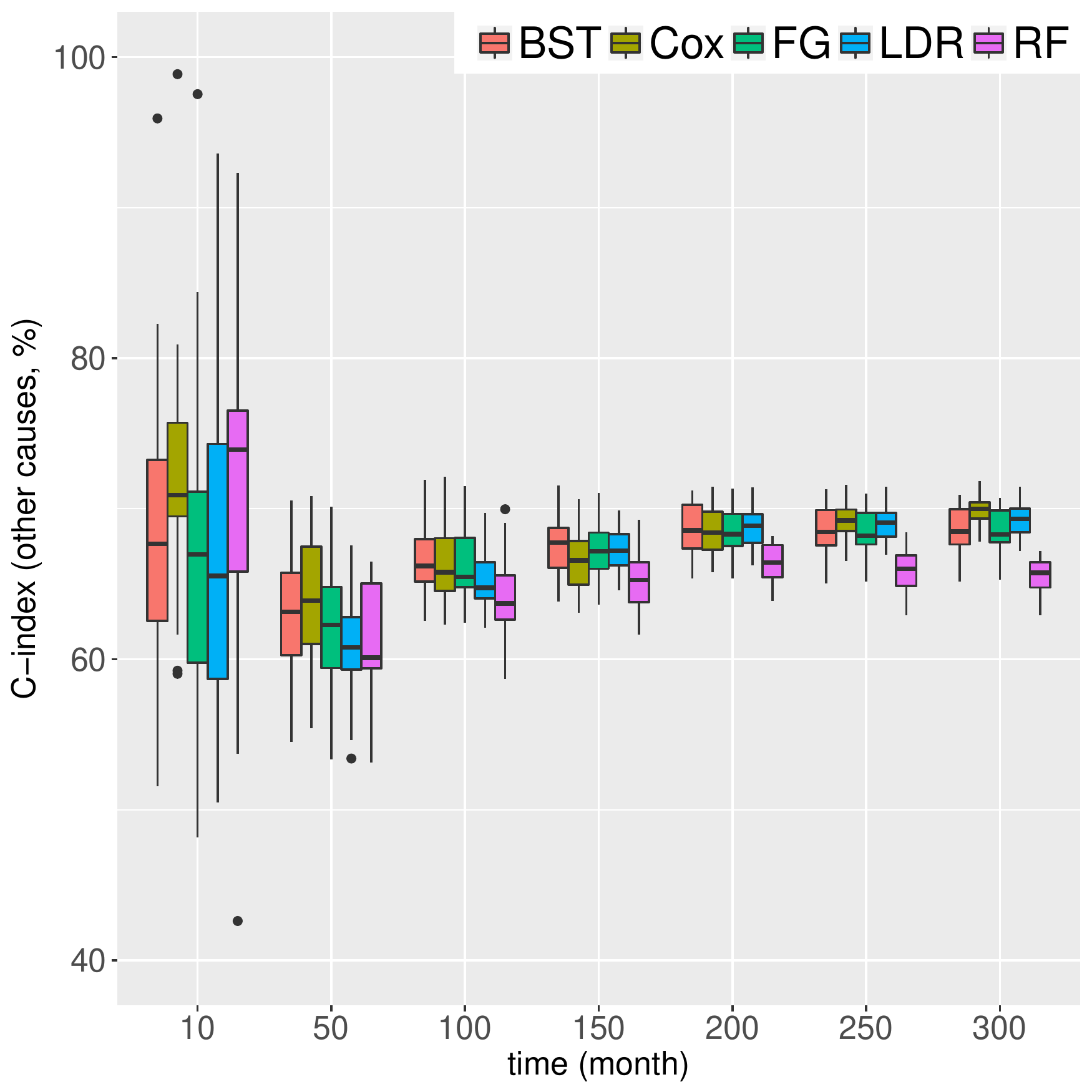}\vspace{-3mm}\label{nomissing_2}
 \caption{OC.}\vspace{-2.5mm}
 \end{subfigure}%
 ~
 \begin{subfigure}[t]{0.24\textwidth}
 \centering
\includegraphics[width=1\linewidth]{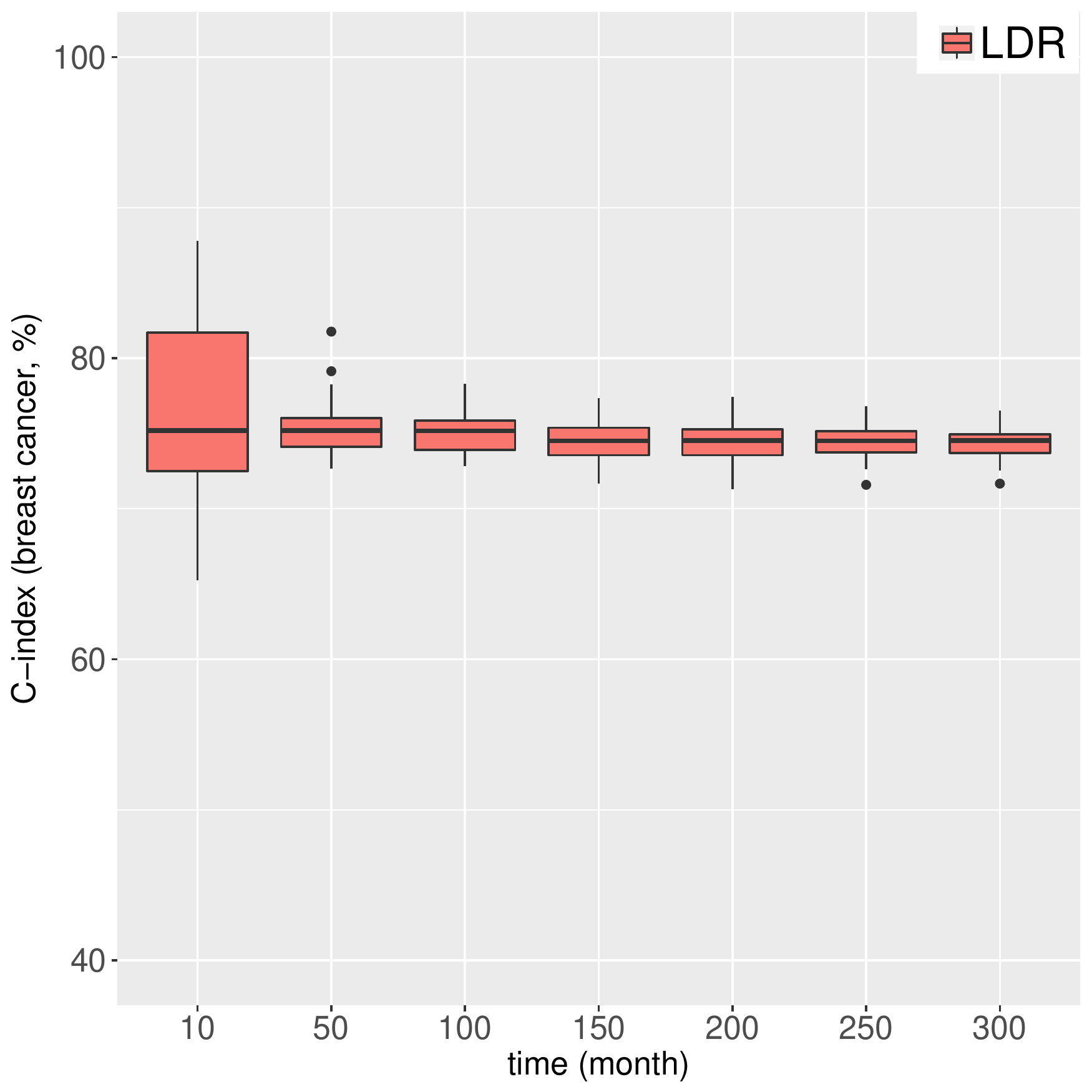}\vspace{-3mm}\label{missing_1}
 \caption{BC, with unknown\\\mbox{ }~~~ event types.}\vspace{-2.5mm}
 \end{subfigure}%
 ~ 
 \begin{subfigure}[t]{0.24\textwidth}
 \centering
\includegraphics[width=1\linewidth]{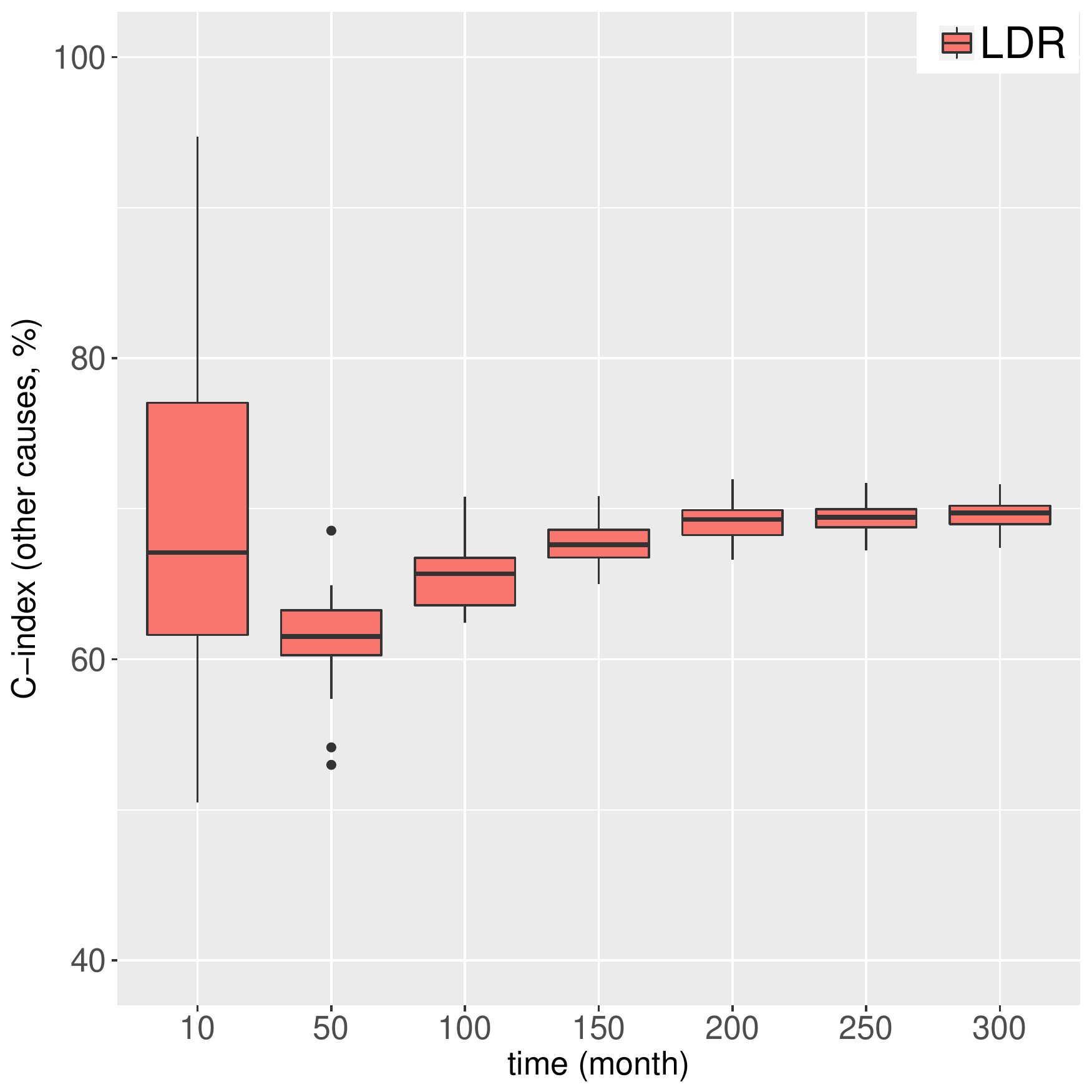}\vspace{-3mm}\label{missing_2}
 \caption{OC, with unknown\\\mbox{ }~~~ event types.}\vspace{-2.5mm}
 \end{subfigure}
\caption{C-indices for SEER breast cancer data.}\vspace{-4mm}\label{cindex_seer}
\end{figure}

We first eliminate all observations with unknown causes of death, so we can make comparison between LDR, Cox, FG, BST, and RF. We take 20 random training/testing partitions of the dataset, in each of which $80\%$ of observations are used as training and the remaining $20\%$ as testing. In Figure \ref{cindex_seer}, panels (a) and (b) show the boxplots of C-indices for BC and OC, respectively, obtained from the 20 testing sets by the five models at months $10, 50, 100, \cdots, 300$. For BC the C-indices by LDR are comparable to those by the other four approaches until month 150 and slightly higher afterwards. For the OC the C-indices by LDR are slightly lower than those by Cox, FG, and BST, but become similar after month 100. Also note that RF underperforms the other four approaches since month 100 for BC and month 50 for OC. Comparable C-indices from LDR, Cox, FG, and BST imply monotonic impacts of covariates on survival times under both risks. In fact, for either risk we learn a sub-risk which dominates the others in terms of weights. Furthermore, we analyze the SEER dataset by LDR using the same training/testing partitions, but additionally including the observations having missing causes of death into the 20 training sets, and show the testing C-indices in panels (c) and (d) of Figure \ref{cindex_seer}. We see the testing C-indices are very similar to those in (a) and (b). More importantly, LDR provides a probabilistic inference on missing time to event or missing causes during the model training procedure.

In Appendix E we further provide the Brier scores \cite{gerds2008performance,steyerberg2010assessing} of each risk in all data sets over time. Brier score quantifies the deviation of predicted CIF's from the actual outcomes and a smaller value implies a better model performance \cite{van2011dynamic}. Tables \ref{tab:linear_risk1}-\ref{tab:seer_risk2} in Appendix E show Brier scores by the models compared on the four data sets, indicating the model out-of-sample prediction performance is basically consistent with those quantified by C-indices. Specifically, for the cases of synthetic data 1, {{SEER, and both ABC and GCB of DLBCL}}, where C-indices imply linear covariate effects, the Brier scores are comparable for Cox, FG, BST, and LDR, and slightly smaller than those of RF. For synthetic data 2 and T3 of DLBCL where C-indices imply nonlinear covariate effects, the Brier scores by LDR and RF are smaller than those by Cox, FG, and BST. Moreover, the Brier scores by LDR are slightly larger than those of RF for synthetic data 2 but smaller for T3 of DLBCL. 

\begin{figure}[!t] 
 \centering
 \begin{subfigure}[t]{0.36\textwidth} 
 \centering
\includegraphics[width=1\linewidth]{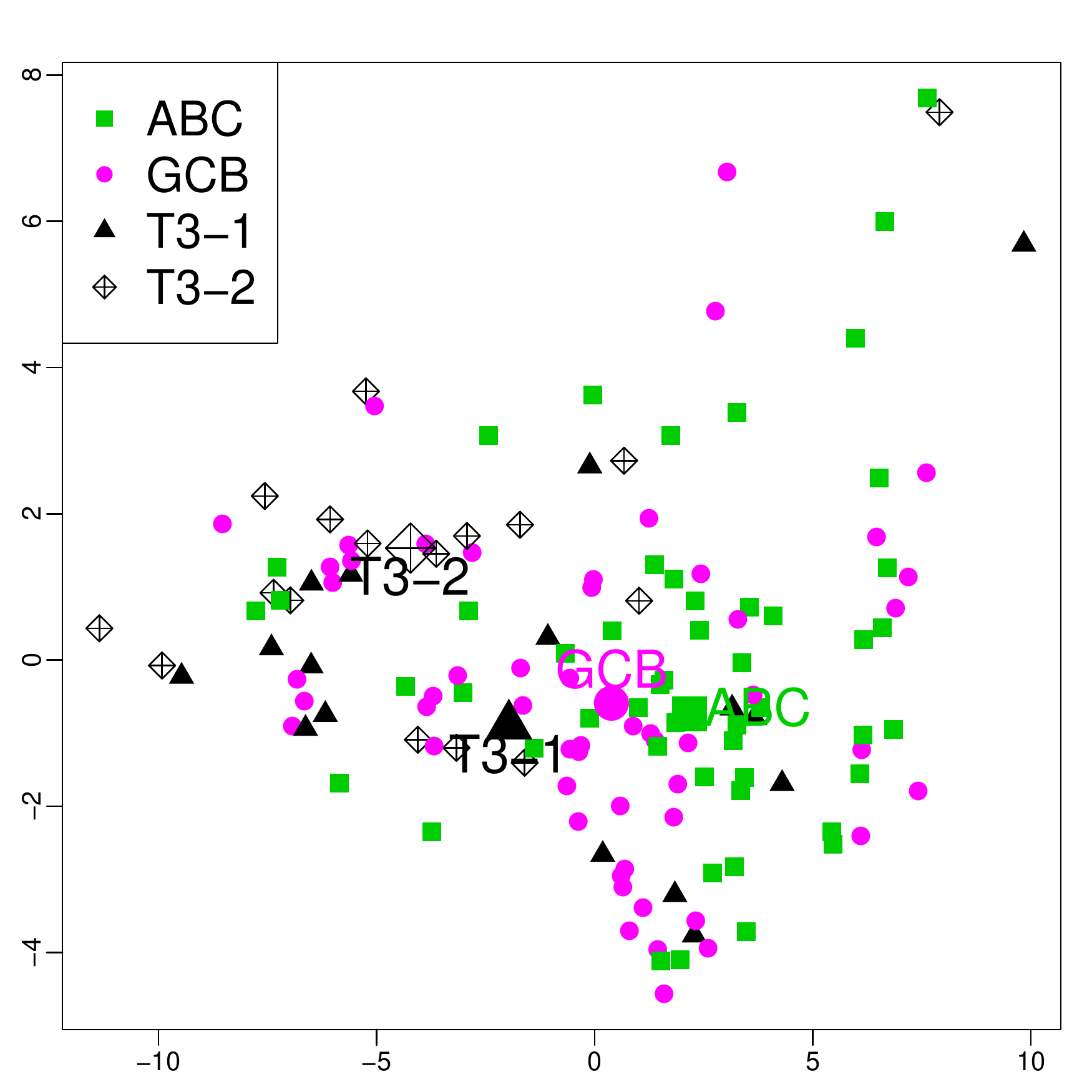}\vspace{-4mm}\label{iso_dlbcl}
 \caption{DLBCL.}\vspace{-4mm}
 \end{subfigure}%
 ~ 
 \begin{subfigure}[t]{0.36\textwidth} 
 \centering
\includegraphics[width=1\linewidth]{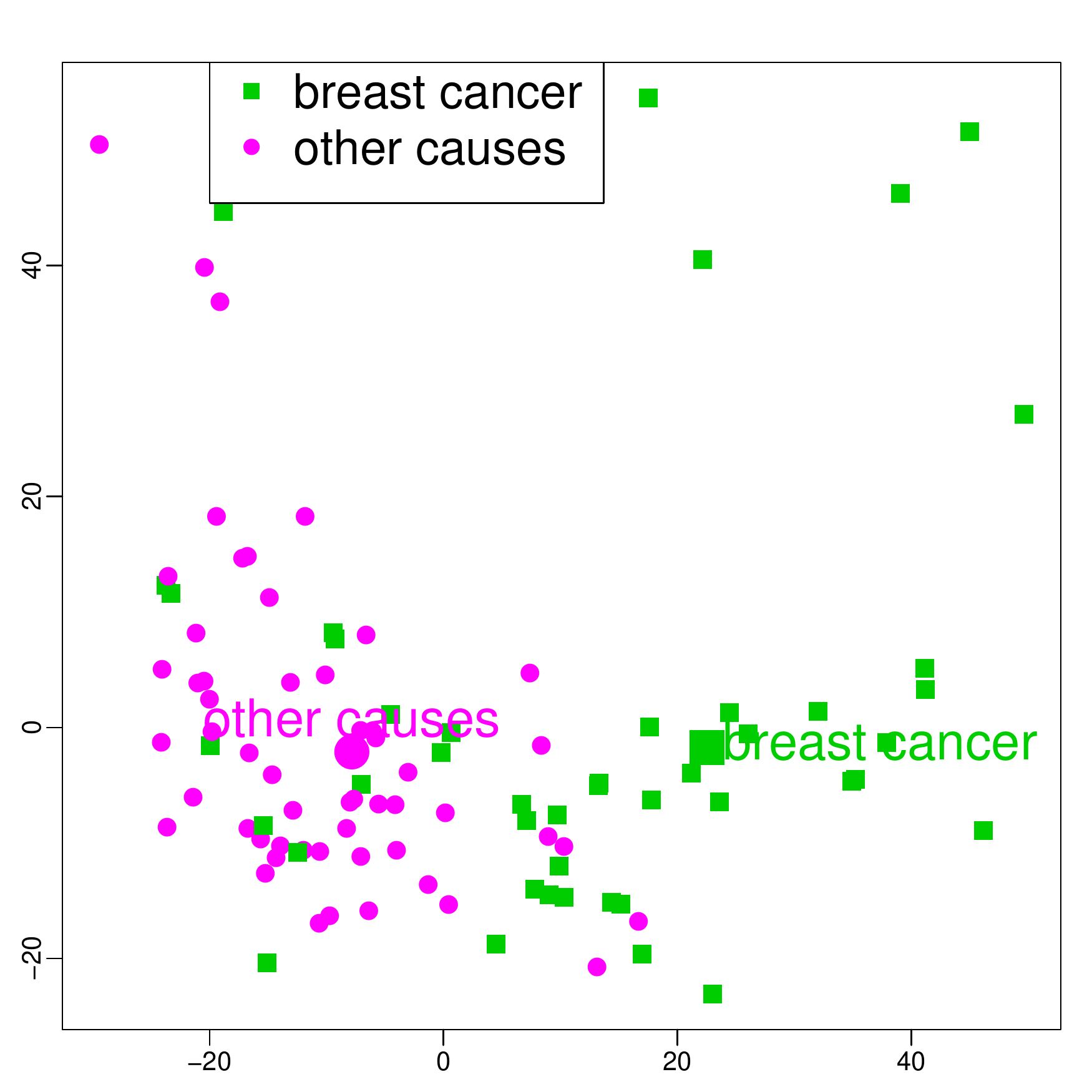}\vspace{-4mm}\label{cosh_risk2}
 \caption{SEER.}
 \end{subfigure}%
\caption{Isomap visualization of the observations and inferred sub-risk representatives.} 
\label{isomap}
\end{figure}

To show the interpretability of LDR, we visualize representative individuals, each of which suffers from an inferred sub-risk. 
Specifically, for each inferred sub-risk $k$ under risk $j$, we find the representative by evaluating a weighted average of all uncensored observations as $\sum_i w_{ijk}\xv_i / \sum_i w_{ijk}$, where $w_{ijk}=\mathbb{E}\left(\frac{\lambda_{ijk}}{
\sum_{j'}\sum_{k'} \lambda_{ij'k'}
}\right)$, $\lambda_{ijk}\sim \mbox{Gamma}(\hat r_{jk},e^{\xv_i'\hat\betav_{jk}})$, and $\hat r_{jk}$ and $\hat\betav_{jk}$ are the estimated values 
 of $r_{jk}$ and $\betav_{jk}$, respectively. The weight $w_{ijk}$ extracts the component of $\xv_i$ that is likely to make the event of sub-risk $k$ under risk $j$ first occur. Then we implement an Isomap algorithm \cite{tenenbaum2000global} and visualize in Figure \ref{isomap} the representatives along with uncensored observations in both DLBCL and SEER. Details are provided in the Appendix.

In Figure \ref{isomap}, small symbols denote uncensored observations and large ones the representatives. Panels (a) and (b) show the representatives suffering from sub-risks in the DLBCL and SEER dataset, respectively. In panel (a), we use green for ABC, pink for GCB, and black for T3. The only representative suffering from ABC (GCB) is surrounded by small green (pink) symbols, indicating they signify a typical gene expression profile that may result in the corresponding malignant transformation. There are two representatives suffering from the two sub-risks of T3, denoted by a large triangle and a large diamond, respectively. They approximately lie in the center of the respective cluster of small triangles and diamonds, which denote patients suffering from the corresponding sub-risks of T3 with an estimated probability greater than $0.5$. The two sub-risks of T3 and the representatives verify the heterogeneity of gene expressions under this risk, and strengthen the belief that T3 consists of more than one type of DLBCL \cite{rosenwald2002use}. For the SEER data, we randomly select 100 of the 2088 uncensored observations with known event types for visualization. In panel (b), we use green for BC and pink for OC. LDR learns only one sub-risk for each of these two risks, and place for each risk a representative approximately at the center of the cluster of patients who died of that risk. 

\section{Conclussion}\label{sec_conclusion} 
We propose 
 Lomax delegate racing (LDR) for survival analysis with competing risks. LDR models the survival times under risks as a two-phase race of sub-risks, which not only intuitively explains the mechanism of surviving under competing risks, but also helps model non-monotonic covariate effects. We use the gamma process to support a potentially countably infinite number of sub-risks for each risk, and rely on its inherent shrinkage mechanism to remove unneeded model capacity, making LDR be capable of detecting unknown event subtypes without pre-specifying their numbers. LDR admits a hierarchical representation that facilities the derivation of Gibbs sampling under data augmentation, which can be adapted for various practical situations such as missing event times or types. 
A more scalable (stochastic) gradient descent based maximum a posteriori inference algorithm is also developed for big data applications. Experimental results show that with strong interpretability and outstanding performance, the proposed LDR survival model is an attractive alternative to 
existing ones for various tasks in survival analysis with competing risks.

\subsubsection*{Acknowledgments}
The authors acknowledge the support of Award IIS-1812699 from the U.S. National Science Foundation, and the computational support of Texas Advanced Computing Center.

\bibliographystyle{ieeetr.bst}
\small
\bibliography{References112016.bib,reference.bib,survival_ref.bib}

\begin{thebibliography}{10}

\bibitem{kaplan1958nonparametric}
E.~L. Kaplan and P.~Meier, ``Nonparametric estimation from incomplete
  observations,'' {\em Journal of the American statistical association},
  vol.~53, no.~282, pp.~457--481, 1958.

\bibitem{wang2016isotonic}
Y.~Wang, B.~Xie, N.~Du, and L.~Song, ``Isotonic {H}awkes processes,'' in {\em
  International conference on machine learning}, pp.~2226--2234, 2016.

\bibitem{mei2017neural}
H.~Mei and J.~M. Eisner, ``The neural {H}awkes process: {A} neurally
  self-modulating multivariate point process,'' in {\em NIPS}, pp.~6754--6764,
  2017.

\bibitem{cox1992regression}
D.~R. Cox, ``Regression models and life-tables,'' in {\em Breakthroughs in
  statistics}, pp.~527--541, Springer, 1992.

\bibitem{ballester1997mortality}
F.~Ballester, D.~Corella, S.~P{\'e}rez-Hoyos, M.~S{\'a}ez, and A.~Herv{\'a}s,
  ``Mortality as a function of temperature. {A} study in {Valencia, Spain},
  1991-1993.,'' {\em International journal of epidemiology}, vol.~26, no.~3,
  pp.~551--561, 1997.

\bibitem{flegal2007cause}
K.~M. Flegal, B.~I. Graubard, D.~F. Williamson, and M.~H. Gail,
  ``Cause-specific excess deaths associated with underweight, overweight, and
  obesity,'' {\em Jama}, vol.~298, no.~17, pp.~2028--2037, 2007.

\bibitem{li2005boosting}
H.~Li and Y.~Luan, ``Boosting proportional hazards models using smoothing
  splines, with applications to high-dimensional microarray data,'' {\em
  Bioinformatics}, vol.~21, no.~10, pp.~2403--2409, 2005.

\bibitem{lu2008boosting}
W.~Lu and L.~Li, ``Boosting method for nonlinear transformation models with
  censored survival data,'' {\em Biostatistics}, vol.~9, no.~4, pp.~658--667,
  2008.

\bibitem{ranganath2016deep}
R.~Ranganath, A.~Perotte, N.~Elhadad, and D.~Blei, ``Deep survival analysis,''
  in {\em Machine Learning for Healthcare Conference}, pp.~101--114, 2016.

\bibitem{katzman2018deepsurv}
J.~L. Katzman, U.~Shaham, A.~Cloninger, J.~Bates, T.~Jiang, and Y.~Kluger,
  ``{DeepSurv: Personalized} treatment recommender system using a {C}ox
  proportional hazards deep neural network,'' {\em BMC medical research
  methodology}, vol.~18, no.~1, p.~24, 2018.

\bibitem{zhu2016deep}
X.~Zhu, J.~Yao, and J.~Huang, ``Deep convolutional neural network for survival
  analysis with pathological images,'' in {\em Bioinformatics and Biomedicine
  (BIBM), 2016 IEEE International Conference on}, pp.~544--547, IEEE, 2016.

\bibitem{chapfuwa2018adversarial}
P.~Chapfuwa, C.~Tao, C.~Li, C.~Page, B.~Goldstein, L.~Carin, and R.~Henao,
  ``Adversarial time-to-event modeling,'' in {\em ICML}, 2018.

\bibitem{fernandez2016gaussian}
T.~Fern{\'a}ndez, N.~Rivera, and Y.~W. Teh, ``Gaussian processes for survival
  analysis,'' in {\em NIPS}, pp.~5021--5029, 2016.

\bibitem{luck2017deep}
M.~Luck, T.~Sylvain, H.~Cardinal, A.~Lodi, and Y.~Bengio, ``Deep learning for
  patient-specific kidney graft survival analysis,'' {\em arXiv preprint
  arXiv:1705.10245}, 2017.

\bibitem{li2016multi}
Y.~Li, J.~Wang, J.~Ye, and C.~K. Reddy, ``A multi-task learning formulation for
  survival analysis,'' in {\em Proceedings of the 22nd ACM SIGKDD International
  Conference on Knowledge Discovery and Data Mining}, pp.~1715--1724, ACM,
  2016.

\bibitem{yu2011learning}
C.-N. Yu, R.~Greiner, H.-C. Lin, and V.~Baracos, ``Learning patient-specific
  cancer survival distributions as a sequence of dependent regressors,'' in
  {\em NIPS}, pp.~1845--1853, 2011.

\bibitem{miscouridoudeep}
X.~Miscouridou, A.~Perotte, N.~Elhadad, and R.~Ranganath, ``Deep survival
  analysis: Nonparametrics and missingness,'' in {\em Machine Learning for
  Healthcare Conference}, 2018.

\bibitem{kalbfleisch2011statistical}
J.~D. Kalbfleisch and R.~L. Prentice, {\em The statistical analysis of failure
  time data}, vol.~360.
\newblock John Wiley \& Sons, 2011.

\bibitem{austin2016introduction}
P.~C. Austin, D.~S. Lee, and J.~P. Fine, ``Introduction to the analysis of
  survival data in the presence of competing risks,'' {\em Circulation},
  vol.~133, no.~6, pp.~601--609, 2016.

\bibitem{putter2007tutorial}
H.~Putter, M.~Fiocco, and R.~B. Geskus, ``Tutorial in biostatistics:
  {C}ompeting risks and multi-state models,'' {\em Statistics in medicine},
  vol.~26, no.~11, pp.~2389--2430, 2007.

\bibitem{lau2009competing}
B.~Lau, S.~R. Cole, and S.~J. Gange, ``Competing risk regression models for
  epidemiologic data,'' {\em American journal of epidemiology}, vol.~170,
  no.~2, pp.~244--256, 2009.

\bibitem{fine1999proportional}
J.~P. Fine and R.~J. Gray, ``A proportional hazards model for the
  subdistribution of a competing risk,'' {\em Journal of the American
  statistical association}, vol.~94, no.~446, pp.~496--509, 1999.

\bibitem{wolbers2014concordance}
M.~Wolbers, P.~Blanche, M.~T. Koller, J.~C. Witteman, and T.~A. Gerds,
  ``Concordance for prognostic models with competing risks,'' {\em
  Biostatistics}, vol.~15, no.~3, pp.~526--539, 2014.

\bibitem{ishwaran2014random}
H.~Ishwaran, T.~A. Gerds, U.~B. Kogalur, R.~D. Moore, S.~J. Gange, and B.~M.
  Lau, ``Random survival forests for competing risks,'' {\em Biostatistics},
  vol.~15, no.~4, pp.~757--773, 2014.

\bibitem{barrett2013gaussian}
J.~E. Barrett and A.~C. Coolen, ``Gaussian process regression for survival data
  with competing risks,'' {\em arXiv preprint arXiv:1312.1591}, 2013.

\bibitem{alaa2009deep}
A.~M. Alaa and M.~v.~d. Schaar, ``Deep multi-task gaussian processes for
  survival analysis with competing risks,'' in {\em NIPS}, 2017.

\bibitem{lee2018deephit}
C.~Lee, W.~R. Zame, J.~Yoon, and M.~van~der Schaar, ``{DeepHit}: A deep
  learning approach to survival analysis with competing risks,'' AAAI, 2018.

\bibitem{crowder2001classical}
M.~J. Crowder, {\em Classical competing risks}.
\newblock CRC Press, 2001.

\bibitem{Ross:2006:IPM:1197141}
S.~M. Ross, {\em Introduction to Probability Models}.
\newblock Academic Press, 10th~ed., 2006.

\bibitem{caron2012bayesian}
F.~Caron and Y.~W. Teh, ``Bayesian nonparametric models for ranked data,'' in
  {\em NIPS}, pp.~1520--1528, 2012.

\bibitem{lomax1954business}
K.~Lomax, ``Business failures: Another example of the analysis of failure
  data,'' {\em Journal of the American Statistical Association}, vol.~49,
  no.~268, pp.~847--852, 1954.

\bibitem{myhre1982screen}
J.~Myhre and S.~Saunders, ``Screen testing and conditional probability of
  survival,'' {\em Lecture Notes-Monograph Series}, pp.~166--178, 1982.

\bibitem{howlader2002bayesian}
H.~A. Howlader and A.~M. Hossain, ``Bayesian survival estimation of {P}areto
  distribution of the second kind based on failure-censored data,'' {\em
  Computational statistics \& data analysis}, vol.~38, no.~3, pp.~301--314,
  2002.

\bibitem{cramer2011progressively}
E.~Cramer and A.~B. Schmiedt, ``Progressively {Type-II} censored competing
  risks data from {L}omax distributions,'' {\em Computational Statistics \&
  Data Analysis}, vol.~55, no.~3, pp.~1285--1303, 2011.

\bibitem{hemmati2017adaptive}
F.~Hemmati and E.~Khorram, ``On adaptive progressively {Type-II} censored
  competing risks data,'' {\em Communications in Statistics-Simulation and
  Computation}, pp.~1--23, 2017.

\bibitem{al2001statistical}
S.~Al-Awadhi and M.~Ghitany, ``Statistical properties of {Poisson-Lomax}
  distribution and its application to repeated accidents data,'' {\em Journal
  of Applied Statistical Science}, vol.~10, no.~4, pp.~365--372, 2001.

\bibitem{childs2001order}
A.~Childs, N.~Balakrishnan, and M.~Moshref, ``Order statistics from
  non-identical right-truncated {L}omax random variables with applications,''
  {\em Statistical Papers}, vol.~42, no.~2, pp.~187--206, 2001.

\bibitem{giles2013bias}
D.~E. Giles, H.~Feng, and R.~T. Godwin, ``On the bias of the maximum likelihood
  estimator for the two-parameter {Lomax} distribution,'' {\em Communications
  in Statistics-Theory and Methods}, vol.~42, no.~11, pp.~1934--1950, 2013.

\bibitem{varma2014prevalence}
R.~Varma, N.~M. Bressler, Q.~V. Doan, M.~Gleeson, M.~Danese, J.~K. Bower,
  E.~Selvin, C.~Dolan, J.~Fine, S.~Colman, {\em et~al.}, ``Prevalence of and
  risk factors for diabetic macular edema in the {United States},'' {\em JAMA
  ophthalmology}, vol.~132, no.~11, pp.~1334--1340, 2014.

\bibitem{zhou2016augmentable}
M.~Zhou, Y.~Cong, and B.~Chen, ``Augmentable gamma belief networks,'' {\em
  Journal of Machine Learning Research}, vol.~17, no.~163, pp.~1--44, 2016.

\bibitem{SoftplusReg_2016}
M.~Zhou, ``Softplus regressions and convex polytopes,'' {\em arXiv:1608.06383},
  2016.

\bibitem{ferguson73}
T.~S. Ferguson, ``A {B}ayesian analysis of some nonparametric problems,'' {\em
  Ann. Statist.}, vol.~1, no.~2, pp.~209--230, 1973.

\bibitem{NBP2012}
M.~Zhou and L.~Carin, ``Negative binomial process count and mixture modeling,''
  {\em IEEE Trans. Pattern Anal. Mach. Intell.}, vol.~37, no.~2, pp.~307--320,
  2015.

\bibitem{LGNB_ICML2012}
M.~Zhou, L.~Li, D.~Dunson, and L.~Carin, ``Lognormal and gamma mixed negative
  binomial regression,'' in {\em ICML}, pp.~1343--1350, 2012.

\bibitem{polson2013bayesian}
N.~G. Polson, J.~G. Scott, and J.~Windle, ``Bayesian inference for logistic
  models using {P{\'o}lya--Gamma} latent variables,'' {\em J. Amer. Statist.
  Assoc.}, vol.~108, no.~504, pp.~1339--1349, 2013.

\bibitem{binder2009boosting}
H.~Binder, A.~Allignol, M.~Schumacher, and J.~Beyersmann, ``Boosting for
  high-dimensional time-to-event data with competing risks,'' {\em
  Bioinformatics}, vol.~25, no.~7, pp.~890--896, 2009.

\bibitem{saha2010time}
P.~Saha and P.~Heagerty, ``Time-dependent predictive accuracy in the presence
  of competing risks,'' {\em Biometrics}, vol.~66, no.~4, pp.~999--1011, 2010.

\bibitem{rosenwald2002use}
A.~Rosenwald, G.~Wright, W.~C. Chan, J.~M. Connors, E.~Campo, R.~I. Fisher,
  R.~D. Gascoyne, H.~K. Muller-Hermelink, E.~B. Smeland, J.~M. Giltnane, {\em
  et~al.}, ``The use of molecular profiling to predict survival after
  chemotherapy for diffuse large-{B}-cell lymphoma,'' {\em New England Journal
  of Medicine}, vol.~346, no.~25, pp.~1937--1947, 2002.

\bibitem{seer}
S.~R. P. S. S.~B. National Cancer~Institute, DCCPS, {\em Surveillance,
  Epidemiology, and End Results (SEER) Program Research Data (1973-2014)},
  Released April 2017, based on the November 2016 submission.
\newblock Released April 2017, based on the November 2016 submission.

\bibitem{gerds2008performance}
T.~A. Gerds, T.~Cai, and M.~Schumacher, ``The performance of risk prediction
  models,'' {\em Biometrical Journal: Journal of Mathematical Methods in
  Biosciences}, vol.~50, no.~4, pp.~457--479, 2008.

\bibitem{steyerberg2010assessing}
E.~W. Steyerberg, A.~J. Vickers, N.~R. Cook, T.~Gerds, M.~Gonen, N.~Obuchowski,
  M.~J. Pencina, and M.~W. Kattan, ``Assessing the performance of prediction
  models: {A} framework for some traditional and novel measures,'' {\em
  Epidemiology (Cambridge, Mass.)}, vol.~21, no.~1, p.~128, 2010.

\bibitem{van2011dynamic}
H.~Van~Houwelingen and H.~Putter, {\em Dynamic prediction in clinical survival
  analysis}.
\newblock CRC Press, 2011.

\bibitem{tenenbaum2000global}
J.~B. Tenenbaum, V.~De~Silva, and J.~C. Langford, ``A global geometric
  framework for nonlinear dimensionality reduction,'' {\em Science}, vol.~290,
  no.~5500, pp.~2319--2323, 2000.

\bibitem{moschopoulos1985distribution}
P.~G. Moschopoulos, ``The distribution of the sum of independent gamma random
  variables,'' {\em Annals of the Institute of Statistical Mathematics},
  vol.~37, no.~1, pp.~541--544, 1985.

\bibitem{pec}
T.~A. Gerds, {\em pec: Prediction Error Curves for Risk Prediction Models in
  Survival Analysis}, 2017.
\newblock R package version 2.5.4.

\bibitem{riskRegression}
T.~A. Gerds and T.~H. Scheike, {\em riskRegression: Risk Regression Models for
  Survival Analysis with Competing Risks}, 2015.
\newblock R package version 1.1.7.

\bibitem{cmprsk}
B.~Gray, {\em cmprsk: Subdistribution Analysis of Competing Risks}, 2014.
\newblock R package version 2.2-7.

\bibitem{coxboost}
H.~Binder, {\em CoxBoost: Cox models by likelihood based boosting for a single
  survival endpoint or competing risks}, 2013.
\newblock R package version 1.4.

\bibitem{RSF}
H.~Ishwaran and U.~Kogalur, {\em Random Forests for Survival, Regression, and
  Classification (RF-SRC)}, 2018.
\newblock R package version 2.6.0.

\bibitem{cormen2009introduction}
T.~H. Cormen, {\em Introduction to algorithms}.
\newblock MIT press, 2009.

\bibitem{borg2005modern}
I.~Borg and P.~J. Groenen, {\em Modern multidimensional scaling: Theory and
  applications}.
\newblock Springer Science \& Business Media, 2005.

\end{thebibliography}
\normalsize
\newpage
\appendix
\begin{center}
\Large{\textbf{Nonparametric Bayesian Lomax delegate racing for\\ survival analysis with competing risks: Appendix}}\\\vspace{2mm}
\normalsize Quan Zhang and Mingyuan Zhou

\end{center}
\section{Marginal distribution of failure time in LDR}
\begin{theorem}
If $t_i\sim \mbox{Gamma}(1, 1/\lambda_{i\bullet\bullet})$ with $\lambda_{i\bullet\bullet}=\sum_{j,k}\lambda_{ijk}$ and $\lambda_{ijk}\sim \mbox{Gamma}(r_{jk}, 1/b_{ijk})$, the PDF of $t_i$ given $\{r_{jk}\}$ and $\{b_{ijk}\}$ is 
\begin{align*}
f(t_i\given \{r_{jk}\}_{j,k},\{b_{ijk}\}_{j,k})=c_i\sum_{m=0}^\infty\frac{(\rho_i+m) \delta_{im}b_{i(1)}^{\rho_i+m}}{(t_i+b_{i(1)})^{1+\rho_i+m}},
\end{align*}
and the cumulative density function (CDF) is
\begin{align}
P(t_i<q\given \{r_{jk}\}_{j,k},\{b_{ijk}\}_{j,k})=1-c_i\sum_{m=0}^\infty\frac{\delta_{im}b_{i(1)}^{\rho_i+m}}{(q+b_{i(1)})^{\rho_i+m}},\label{cdf_convolution}
\end{align}
where $c_{i}=\prod_{j,k}\left(\frac{b_{ijk}}{b_{i(1)}}\right)^{r_{jk}}$, $b_{i(1)}=\max_{j,k} b_{ijk}$, $\rho_i=\sum_{j,k}r_{jk}$, $\delta_{i0}=1$, $\delta_{im+1}=\frac{1}{m+1}\sum_{h=1}^{m+1}h\gamma_{ih}\delta_{im+1-h}$ for $m\geq 1$, and $\gamma_{ih}=\sum_{j,k}\frac{r_{jk}}{h}\left(1-\frac{b_{ijk}}{b_{i(1)}}\right)^h$.
\end{theorem}
It is difficult to utilize the PDF or CDF of $t_i$ in the form of series, but we can use a finite truncation to approximate (\ref{cdf_convolution}). Concretely, as $P(t_i<\infty\given n_i=1, \{r_{jk}\}_{j,k},\{b_{ijk}\}_{j,k})=c_i\sum_{m=0}^\infty\delta_{im}=1$, we find an $M$ so large that $c_i\sum_{m=0}^M\delta_{im}$ close to $1$ (say no less than $0.9999$), and use $1-c_i\sum_{m=0}^M\frac{\delta_{im}b_{i(1)}^{\rho_i+m}}{(q+b_{i(1)})^{\rho_i+m}}$ as an approximation. Consequently, sampling $t_i$ is feasible by inverting the approximated CDF for general cases. We have tried prediction by finite truncation on some synthetic data and found $M$ is mostly between 10 and 30 which is computationally acceptable. 
\begin{proof}
We first study the distribution of gamma convolution. Specifically, if $\lambda_t\overset{ind}{\sim} \mbox{Gamma}(r_t, 1/b_t)$ with $r_t,b_t\in \mathbb R_+$, then the PDF of $\lambda=\sum_{t=1}^T$ can be written in a form of series \cite{moschopoulos1985distribution} as 
\begin{align*}
f(\lambda\given r_1, b_1,\cdots,r_T, b_T)=\begin{cases}
c\sum_{m=0}^\infty\frac{\delta_{m} \lambda^{\rho+m-1}e^{-\lambda b_{(1)}}}{\Gamma(\rho+m)/b_{(1)}^{\rho+m}} &\mbox{ if } \lambda>0,\\
0 &\mbox{ otherwise}, 
\end{cases}
\end{align*} 
where $c=\prod_{t=1}^T\left(\frac{b_t}{b_{(1)}}\right)^{r_t}$, $b_{(1)}=\max_t b_t$, $\rho=\sum_{t=1}^T r_t$, $\delta_{0}=1$, $\delta_{m+1}=\frac{1}{m+1}\sum_{h=1}^{m+1} h \gamma_{h} \delta_{m+1-h}$ and $\gamma_{h}=\sum_{t=1}^T r_{t}\left(1-\frac{b_{t}}{b_{(1)}}\right)^{h}/h$. 
\cite{moschopoulos1985distribution} proved that $0<\gamma_{ih}\leq \rho_i b_{i0}^h/h$ and $0<\delta_{im}\leq \frac{\Gamma(\rho_i+m)b_{i0}^m}{\Gamma(\rho_i)m!}$ where $b_{i0}=max_{j,k}(1-\frac{b_{ijk}}{b_{i(1)}})$. With $n_i\equiv 1$, we want to show the PDF of $t_i$,
\begin{align}
&f(t_i\given \{r_{jk}\}_{j,k},\{b_{ijk}\}_{j,k})\nonumber\\
=&\int_0^{\infty} f(t_i\given \lambda_{i\bullet\bullet})f(\lambda_{i\bullet\bullet}\given \{r_{jk}\}_{j,k},\{b_{ijk}\}_{j,k}) d\lambda_{i\bullet\bullet}\nonumber\\
=&\int_0^{\infty}\sum_{m=0}^\infty \frac{c_i\delta_{im}t_i^{n_i-1} \lambda_{i\bullet\bullet}^{n_i+\rho_i+m-1}\exp (-t_i\lambda_{i\bullet\bullet}-b_{i(1)}\lambda_{i\bullet\bullet})}{\Gamma(n_i)\Gamma(\rho_i+m)} d\lambda_{i\bullet\bullet} \nonumber\\
=&\sum_{m=0}^\infty \int_0^{\infty}\frac{c_i\delta_{im}t_i^{n_i-1} \lambda_{i\bullet\bullet}^{n_i+\rho_i+m-1}\exp (-t_i\lambda_{i\bullet\bullet}-b_{i(1)}\lambda_{i\bullet\bullet})}{\Gamma(n_i)\Gamma(\rho_i+m)} d\lambda_{i\bullet\bullet} \label{pdf_interchange}\\
=&\frac{c_i t_i^{n_i-1}}{\Gamma(n_i)}\sum_{m=0}^\infty\frac{\Gamma(n_i+\rho_i+m) \delta_{im}b_{i(1)}^{\rho_i+m}}{\Gamma(\rho_i+m)(t_i+b_{i(1)})^{n_i+\rho_i+m}},\nonumber
\end{align}
which suffices to prove the equality in \eqref{pdf_interchange}. Note that
\begin{align*}
&f(t_i\given n_i,\lambda_{i\bullet\bullet})f(\lambda_{i\bullet\bullet}\given \{r_{jk}\}_{j,k},\{b_{ijk}\}_{j,k})\\
=& \frac{c_i}{\Gamma(n_i)}t_i^{n_i-1}\lambda_{i\bullet\bullet}^{n_i+\rho_i-1} b_{i(1)}^{\rho_i}\exp (-t_i\lambda_{i\bullet\bullet}-b_{i(1)}\lambda_{i\bullet\bullet})\sum_{m=0}^\infty \frac{\Gamma(\rho_i+m)}{\delta_{im}b_{i(1)}^m\lambda_{i\bullet\bullet}^m}\\
\leq &\frac{c_i}{\Gamma(n_i)}t_i^{n_i-1}\lambda_{i\bullet\bullet}^{n_i+\rho_i-1} b_{i(1)}^{\rho_i}\exp (-t_i\lambda_{i\bullet\bullet}-b_{i(1)}\lambda_{i\bullet\bullet}) \sum_{m=0}^\infty \frac{(b_{i0}b_{i(1)}\lambda_{i\bullet\bullet})^m}{\Gamma(\rho_i)m!}\\
=&\frac{c_i}{\Gamma(n_i)}t_i^{n_i-1}\lambda_{i\bullet\bullet}^{n_i+\rho_i-1} b_{i(1)}^{\rho_i} \exp (-t_i\lambda_{i\bullet\bullet}-b_{i(1)}\lambda_{i\bullet\bullet} +b_{i0}b_{i(1)}\lambda_{i\bullet\bullet} ),
\end{align*}
which shows the uniform convergence of $f(t_i\given n_i,\lambda_{i\bullet\bullet})f(\lambda_{i\bullet\bullet}\given \{r_{jk}\}_{j,k},\{b_{ijk}\}_{j,k})$. So the integration and countable summation are interchangeable, and consequently, \eqref{pdf_interchange} holds. Next we want to show the CDF of $t_i$,
\begin{align}
P(t_i<q\given n_i, \{r_{jk}\}_{j,k},\{b_{ijk}\}_{j,k})=&\int_0^q \frac{c_i t_i^{n_i-1}}{\Gamma(n_i)}\sum_{m=0}^\infty\frac{\Gamma(n_i+\rho_i+m) \delta_{im}b_{i(1)}^{\rho_i+m}}{\Gamma(\rho_i+m)(t_i+b_{i(1)})^{n_i+\rho_i+m}} dt_i \nonumber\\
=& \sum_{m=0}^\infty \int_0^q \frac{c_i t_i^{n_i-1}}{\Gamma(n_i)}\frac{\Gamma(n_i+\rho_i+m) \delta_{im}b_{i(1)}^{\rho_i+m}}{\Gamma(\rho_i+m)(t_i+b_{i(1)})^{n_i+\rho_i+m}} dt_i. \label{cdf_interchange}
\end{align}
It suffices to show \eqref{cdf_interchange}. Note that
\begin{align*}
&\frac{c_i t_i^{n_i-1}}{\Gamma(n_i)}\sum_{m=0}^\infty\frac{\Gamma(n_i+\rho_i+m) \delta_{im}b_{i(1)}^{\rho_i+m}}{\Gamma(\rho_i+m)(t_i+b_{i(1)})^{n_i+\rho_i+m}}\\
\leq &\frac{c_i t_i^{n_i-1}}{\Gamma(n_i)} \sum_{m=0}^\infty \frac{\Gamma(n_i+\rho_i+m)b_{i(1)}^{\rho_i+m}\Gamma(n_i+\rho_i+m)}{\Gamma(\rho_i+m)(t_i+b_{i(1)})^{n_i+\rho_i+m}\Gamma(\rho_i)m!}\\
=&\frac{c_i t_i^{n_i-1}}{\Gamma(n_i)}\frac{\Gamma(\rho_i+n_i)b_{i(1)}^{\rho_i} }{\Gamma(\rho_i)(t_i+b_{i(1)})^{n_i+\rho_i} } \sum_{m=0}^\infty\left[\frac{\Gamma(n_i+\rho_i+m)}{\Gamma_{n_i+\rho_i}m!}\left(\frac{b_{i(1)}}{t_i+b_{i(1)}}\right)^m \right]\\
=&\frac{c_i t_i^{n_i-1} \Gamma(\rho_i+n_i)b_{i(1)}^{\rho_i}t_i^{n_i+\rho_i} }{\Gamma(n_i)\Gamma(\rho_i)(t_i+b_{i(1)})^{2(n_i+\rho_i)}}.
\end{align*}
The last equation holds because the summation of a negative binomial probability mass function is 1. So $f(t_i\given n_i, \{r_{jk}\}_{j,k},\{b_{ijk}\}_{j,k})$ is uniformly convergent and \eqref{cdf_interchange} holds. Plugging in $n_i=1$ and calculating the integration, we obtain the CDF of $t_i$.
\end{proof}

\section{Bayesian inference of LDR}\label{sec:mcmc_sa}
With $\xv_i$ denoting the covariates, $y_i$ event type, and $t_i$ the time to event 
of observation $i$, we express the full hierarchical form of LDR defined in \eqref{eq:argmin_argmin}, as 
\begin{align}
&t_i= t_{iy_i},~y_i=\mathop{\mathrm{argmin}}\limits_{j\in\{1,\ldots,J\}} t_{ij},~ 
t_{ij}= t_{ij\kappa_{ij}},~\kappa_{ij}= \mathop{\mathrm{argmin}}\limits_{k\in\{0,\ldots,K\}} t_{ijk},
 \nonumber\\ 
&t_{ijk} \sim \mbox{Exp}
(\lambda_{ijk}),~\lambda_{ijk} \sim \mbox{Gamma}(r_{jk}, e^{\xv_{i}'\betav_{jk}}), ~k=1,\cdots, K,\nonumber\\
&\betav_{jk}\sim \prod_{g=1}^{P} \mathcal{N}(0,\alpha_{gjk}^{-1}), ~
\alpha_{gjk}\sim \mbox{Gamma}(a_0,1/b_0),
 ~r_{jk}\sim \mbox{Gamma}(\gamma_{0j}/K, 1/c_{0j})\notag,
\end{align}
where $k=1,\cdots, K$, $i=1,\cdots, n$, and $ j=1,\cdots,J$. We further let $\gamma_{0j}\sim \mbox{Gamma}(e_0,1/f_0)$, $c_{0j}\sim \mbox{Gamma}(e_1,1/f_1)$, $r_{0}\sim \mbox{Gamma}(e_0,1/f_0)$, and 
set $e_0=f_0=e_1=f_1=0.01$. Let us denote $T_i$ and $T_{ic}$ as the observed failure time and right censoring time, respectively, for observation $i$. Since left censoring is uncommon and not shown in the real datasets analyzed, we only consider right censoring in our inference and leave to readers other types of censoring which can be analogously done. A Gibbs sampler accommodating missing event times or missing event types proceeds by iterating the following steps.

\begin{enumerate} 
\item If $y_i$ is observed, we first sample $\kappa_{iy_i}$ by
$$P(\kappa_{iy_i}=k\given y_i,\cdots)=\frac{\lambda_{iy_ik}}{\sum_{k'=1}^K \lambda_{iy_ik'}}.$$
If $y_i$ is unobserved which means a missing event type, we sample $(y_i,\kappa_{iy_i})$ by 
$$P(y_i=j, \kappa_{iy_i}=k\given \cdots)=\frac{\lambda_{ijk}}{\sum_{j'=1}^S \sum_{k'=1}^K \lambda_{ij'k'}}.$$
We then denote $m_{jk}=\sum_{i:y_i=j} \bm 1 (\kappa_{iy_i}=k)$. Define $n_{ijk}=1$ if $y_i=j$ and $\kappa_{iy_i}=k$, and otherwise $n_{ijk}=0$.
The above sampling procedure means that given the event type $y_i$, we sample the index of the sub-risk that has the minimum survival time.
\item Update $t_{i}$ for $i=1,\cdots,n$, $j=1,\cdots, J$ and $k=1,\cdots,K$.\label{sample_t}
\begin{enumerate}
\item If the failure time $T_i$ is observed, we set $t_i=T_i$. 
\item Otherwise, we let $t_i = T_{ic} + \tilde t_i$, where $(\tilde t_i\given-) \sim 
\mbox{Exp}(\sum_{j=1}^S\sum_{k=1}^K \lambda_{ijk})$ and $T_{ic}$ is the right censoring. Note $T_{ic}=0$ if both event time and censoring time are missing for observation $i$. 
\end{enumerate}

\item Sample ($\lambda_{ijk}\given-)\sim \mbox{Gamma}\left(r_{jk}+n_{ijk},\frac{e^{\xv_i' \betav_{jk}}}{1+t_i e^{\xv_i' \betav_{jk}}} \right)$, for $i=1,\cdots,n$, $j=1,\cdots,J$ and $k=1,\cdots,K$.
\item Sample $\betav_{jk}$, for $j=1,\cdots,J$ and $k=1,\cdots,K$, by P\'olya Gamma (PG) data augmentation.
First Sample $(\omega_{ijk}\given-)\sim \mbox{PG}(r_{jk}+n_{ijk}, \xv_i'\betav_{jk}+\log t_i)$. Then sample $(\betav_{jk}\given-)\sim \mbox{MVN}(\bm \mu_{jk}, \bm\Sigma_{jk})$ where $\bm\Sigma_{jk}=\left(V_{jk}+\bm X'\Omega_{jk}\bm X\right)^{-1}$, $\bm X=[\xv_1',\cdots,\xv_N']'$, $\Omega_{jk}=\mbox{diag}(\omega_{1jk},\cdots,\omega_{njk})$ and $\bm\mu_{jk}=\bm\Sigma_{jk}\left[-\sum_{i=1}^{N}\left(\omega_{ijk}\log t_i +\frac{r_{jk}-n_{ijk}}{2}\right)\xv_i\right]$.
Note to sample from the P\'olya-Gamma distribution, we use a fast and accurate approximate sampler of Zhou \cite{SoftplusReg_2016} that matches the first two moments of the original distribution; we set the truncation level of that sampler as five. 

\item Sample $(\alpha_{vjk}\given-)\sim \mbox{Gamma}\left(a_0+0.5,1/(b_0+0.5\beta_{vjk}^2)\right)$ for $v=0,\cdots,V$, $j=1,\cdots,J$ and $k=1,\cdots,K$.
\item Sample $r_{jk}$ and $\gamma_{0j}$, for $j=1,\cdots,J$ and $k=1,\cdots,K$, by Chinese restaurant table (CRT) data augmentation \cite{NBP2012}.

First sample $(n^{(2)}_{ijk}\given-)\sim \mbox{CRT}(n_{ijk},r_{jk})$, and $(l_{jk}\given-)\sim \mbox{CRT}(\sum_{i=1}^N n^{(2)}_{ijk}, \gamma_{0j}/K)$. Then sample $(r_{jk}\given-)\sim \mbox{Gamma}\left(\sum_{i=1}^N n^{(2)}_{ijk}+\gamma_{0j}/K, \frac{1}{c_{0j}+\sum_{i=1}^N \log(1+t_i e^{\xv_i' \betav_{jk}})} \right)$, and \\
$(\gamma_{0j}\given-)\sim \mbox{Gamma}\left(e_0+\sum_{k=1}^K l_{jk}, \frac{1}{f_0-\frac{1}{K}\sum_{k=1}^K \log(1-p_{jk})}\right)$, where $p_{jk}=\frac{\sum_{i=1}^N \log(1+t_i e^{\xv_i' \betav_{jk}})}{c_{0j}+\sum_{i=1}^N \log(1+t_i e^{\xv_i' \betav_{jk}})}$.
\item Sample $(c_{0j}\given-)\sim \mbox{Gamma}\left(e_1+\gamma_{0j},\frac{1}{f_1+\sum_{k=1}^K r_{jk}}\right)$ for $j=1,\cdots,J$.
\item For $j=1,\cdots,J$ and $k=1,\cdots, K$, prune sub-risk $k$ of risk $j$ for all observations if $m_{jk}=0$, by setting $\lambda_{ijk}\equiv 0$ and $t_{ijk}\equiv \infty$ for $\forall i$. 
\end{enumerate}

\section{Maximum a posteriori estimation}
With the reparameterization that $\lambda_{ijk}=\tilde{\lambda}_{ijk}e^{\bm x_i'\bm\beta_{jk}}$ where $\tilde{\lambda}_{ijk}\overset{iid}{\sim}\mbox{Gamma}(r_{jk},1)$ we first find $p_i$, the likelihood of observation $i$ having event type $y_i$ at event time $t_i$.
\begin{align*}
p_i=\E\left(P(t_i,y_i\given \bm \lambda_i)\right)\equiv \int \left(p_{t_i}\times p_{y_i}\right) p(\tilde {\bm \lambda}_i\given \bm r) d \tilde {\bm \lambda}_i
\end{align*}
where $\tilde{\bm\lambda}_i=\{\tilde{\lambda}_{ijk}\}_{j,k}$,
$p(\tilde {\bm \lambda}_i\given \bm r)=\prod_{j,k}\mbox{Gamma}(r_{jk},1)$, $\bm r=\{r_{jk}\}_{j,k}$, $\mbox{Gamma}(r_{jk},1)$ is the pdf of a gamma distribution with shape $r_{jk}$ and scale $1$, and \\
\begin{align*}
p_{t_i}=& \begin{cases} (\sum_{j,k}\tilde{\lambda}_{ijk}e^{\bm x_i'\bm\beta_{jk}}) \exp\left\{-t_i \sum_{jk} \tilde{\lambda}_{ijk}e^{\bm x_i'\bm\beta_{jk}}\right\} & \mbox{ if } t_i \mbox{ is uncensored and observed,}\\
\exp\left\{-T_{ic} \sum_{jk} \tilde{\lambda}_{ijk}e^{\bm x_i'\bm\beta_{jk}}\right\} & \mbox{ if } t_i \mbox{ is right censored at } T_{ic},\mbox{ i.e., } t_i>T_{ic},\\
1& \mbox{ if } t_i \mbox{ is missing, but } y_i \mbox{ is not,}
\end{cases} \\
p_{y_i}=& \begin{cases} \frac{\sum_k \tilde{\lambda}_{iy_ik}e^{\bm x_i'\bm\beta_{y_ik}}}{\sum_{j,k}\tilde{\lambda}_{ijk}e^{\bm x_i'\bm\beta_{jk}}} & \mbox{ if } y_i \mbox{ is not missing,}\\
1 & \mbox{ if } y_i \mbox{ is missing, but } t_i \mbox{ is not.}
\end{cases}
\end{align*}
Note that we do not define $P(t_i,y_i\given \bm \lambda_i)$ if both $t_i$ and $y_i$ are missing and remove such observations from data. We write $p_{t_i}\equiv p_t(\tilde{\bm\lambda}_i\given \bm r)$ and $p_{y_i}\equiv p_y(\tilde{\bm\lambda}_i\given \bm r)$.

Imposing a prior $p(\bm\beta_{jk})$ on $\bm\beta_{jk}$ and $p(r_{jk})$ on $r_{jk}$, the log posterior is 
\begin{align}
\log P &= \sum_{i} \log p_{i} + \sum_{j,k}\log p(\bm\beta_{jk}) +\sum_{j,k}\log p(r_{jk})+C \label{map_posterior}
\end{align}
where $C$ is a constant function of $\{\bm\beta_{jk}\}$ and $\{r_{jk}\}$. In practice we assume a Student's $t$ distribution with degrees of freedom $a$ on each element of $\bm\beta_{jk}$ and a Gamma$(0.01/K, 1/0.01)$ prior on $r_{jk}$. We also found a Gamma$(1/K, 1)$ prior on $r_{jk}$ or an $l^2$-regularizer, $0.001||\bm r||_2$, is more numerically stable. Then we have 
\begin{align*}
\log P = \sum_{i} \log p_i
+\sum_{v,j,k}-\frac{a+1}{2}\log\left(1+\beta_{vjk}^2/a\right) +\sum_{j,k}\left[(0.01/K-1)\log r_{jk}-0.01 r_{jk}\right]+c
\end{align*}
where $c$ is also a constant function of $\{\bm\beta_{jk}\}$ and $\{r_{jk}\}$. For simplicity, we define $\bm\beta=\{\bm\beta_{jk}\}_{j,k}$. We want to maximize $\log P$ with respect to $\bm\beta$ and $\bm r$. The difficulty lies in $p_i$ being the expectation of $p_{t_i}\times p_{y_i}$ over $\tilde{\bm \lambda}_i$ which is a random variable parameterized by $\bm r$. Now we show how to approximate the derivatives of $\log p_i$ by Monte-Carlo simulation and score function gradients. Specifically,
\begin{align}
\nabla_{\bm\beta}\log p_i=\frac{\int \left[\nabla_{\bm\beta} \left(p_{t_i}\times p_{y_i}\right) \right] p(\tilde {\bm \lambda}_i\given \bm r) d \tilde {\bm \lambda}_i}{\int \left(p_{t_i}\times p_{y_i}\right) p(\tilde {\bm \lambda}_i\given \bm r) d \tilde {\bm \lambda}_i}
\approx \frac{\frac{1}{M}\sum_{m=1}^M \nabla_{\bm\beta}\left[p_t(\tilde{\bm\lambda}_i^{(m)}\given \bm r)\times p_y(\tilde{\bm\lambda}_i^{(m)}\given \bm r) \right] }{\frac{1}{M}\sum_{m=1}^M \left[p_t(\tilde{\bm\lambda}_i^{(m)}\given \bm r)\times p_y(\tilde{\bm\lambda}_i^{(m)}\given \bm r) \right] }\label{eq:partial_beta}
\end{align}
where $M$ is a reasonably large number, say $10$, $\tilde{\bm \lambda}_i^{(m)}=\{\tilde{\lambda}_{ijk}^{(m)}\}_{jk}$ and $\tilde{\lambda}_{ijk}^{(m)}\overset{iid}{\sim}\mbox{Gamma}(r_{jk},1)$, $\forall i=1,\cdots,n$ and $m=1,\cdots,M$. With the fact that $\nabla_{\bm r} p(\tilde {\bm \lambda}_i\given \bm r) =p(\tilde {\bm \lambda}_i\given \bm r) \nabla_{\bm r} \log p(\tilde {\bm \lambda}_i\given \bm r)$, 
\begin{align}
\nabla_{\bm r}\log p_i &=\frac{\int \nabla_{\bm r}\left[ \left(p_{t_i}\times p_{y_i}\right) p(\tilde {\bm \lambda}_i\given \bm r) \right] d \tilde {\bm \lambda}_i}{\int \left(p_{t_i}\times p_{y_i}\right) p(\tilde {\bm \lambda}_i\given \bm r) d \tilde {\bm \lambda}_i}\nonumber\\
&=\frac{\int \left(p_{t_i}\times p_{y_i}\right) \nabla_{\bm r}\log p(\tilde {\bm \lambda}_i\given \bm r) p(\tilde {\bm \lambda}_i\given \bm r) d \tilde {\bm \lambda}_i}{\int \left(p_{t_i}\times p_{y_i}\right) p(\tilde {\bm \lambda}_i\given \bm r) d \tilde {\bm \lambda}_i}\nonumber\\
&\approx \frac{\frac{1}{M}\sum_{m=1}^M p_t(\tilde{\bm\lambda}_i^{(m)}\given \bm r)\times p_y(\tilde{\bm\lambda}_i^{(m)}\given \bm r) \nabla_{\bm r} \log p(\tilde {\bm \lambda}_i^{(m)}\given \bm r) }{\frac{1}{M}\sum_{m=1}^M \left[p_t(\tilde{\bm\lambda}_i^{(m)}\given \bm r)\times p_y(\tilde{\bm\lambda}_i^{(m)}\given \bm r) \right] }\notag\\
&=\sum_{m=1}^M \frac{ p_t(\tilde{\bm\lambda}_i^{(m)}\given \bm r)\times p_y(\tilde{\bm\lambda}_i^{(m)}\given \bm r) } {{\sum_{m'=1}^M \left[p_t(\tilde{\bm\lambda}_i^{(m')}\given \bm r)\times p_y(\tilde{\bm\lambda}_i^{(m')}\given \bm r) \right] }}\nabla_{\bm r} \log p(\tilde {\bm \lambda}_i^{(m)}\given \bm r) .\label{eq:partial_r}
\end{align}

Therefore, we can approximate the derivatives of $\log P$ with respect to $\bm\beta$ and $\bm r$ by plugging in \eqref{eq:partial_beta} and \eqref{eq:partial_r}, respectively, and maximize $-\log P$ by (stochastic) gradient descent.

\section{Description of SEER data and experiment settings}
\subsection{SEER data for survival analysis}
We use breast cancer data from {Surveillance, Epidemiology, and End Results Program} (SEER) of National Cancer Institute between 1973 and 2003. There are two causes of death; the first is breast cancer and the second is \textit{other causes} treated as a whole. Explanatory variables include age of diagnosis, gender, race, marital status, historic stage, behavior type, tumor size, tumor extension, number of malignant tumors, number of regional nodes containing tumor, number of regional nodes that are examined or removed, confirmation type and surgery type. We use dummies for all categorical variables and select a subset of patient collected from the hospital \textit{C503} so that we do not have to consider site effects. We exclude observations with any missing values in explanatory variables. Finally, there are 2647 and 4166 observations in our data if we exclude and include observations with a missing cause of death, respectively.
\subsection{Experiment settings}
We run $10,000$ interations of Gibbs sampler for LDR with the gamma process truncated at $K=10$ for all experiments, take the first $8,000$ as burn-in, and estimate CIF by averaging its estimators from the last $2,000$ iterations. For random survial forests, we set the number of trees equal to $100$ and the number of splits equal to $2$ as suggested by Ishwaran et al. \cite{ishwaran2014random}. We use \texttt{R} for all the analysis: C-indices are estimated by package \texttt{pec} \cite{pec}, the Cox model by \texttt{riskRegression} \cite{riskRegression}, FG by \texttt{cmprsk} \cite{cmprsk}, BST by \texttt{CoxBoost} \cite{coxboost}, and RF by \texttt{randomForestSRC} \cite{RSF}.

Isomap algorithm is often used for nonlinear dimensionality reduction. We first find five nearest neighbors of each observation, and then construct a neighborhood graph where an observation is connected to another with the edge length equal to the Euclidean distance if it is a 5-nearest neighbor. We calculate the shortest path between two nodes of the graph by Floyd–Warshall algorithm \cite{cormen2009introduction} and obtain a geodesic distance matrix with which we compute two-dimensional embeddings by classical multidimensional scaling \cite{borg2005modern}.
 
\section{Additional experimental results}
We first show in Table \ref{tab:linear_risk1} through Table \ref{tab:DLBCL_risk3} the Brier score at the evaluation time for each risk of the synthetic data sets, SEER and DLBCL data, respectively. Brier score (BS) for risk $j$ at time $\tau$ can be estimated by $\mbox{BS}_j(\tau)=\frac{1}{n}\sum_{i=1}^{n}\left[\bm 1(t_i\leq \tau, y_i=j)-P(t_i\leq \tau, y_i=j)\right]^2 $, with a smaller value indicating a better model fit. Note that the model performance quantified by Brier score is basically consistent with quantified by C-indices. For the cases like synthetic data 1, SEER and ABC and GCB of DLBCL, where covariates are believed to be linearly influential by C-indices, the Brier scores are comparable for Cox, FG, BST and LDR, and slightly smaller than those of RF. For synthetic data 2 and T3 of DLBCL where C-indices imply nonlinear covariate effects, the Brier scores of LDR and RF are smaller than those of Cox, FG and BST. Moreover, the Brier score of LDR is slightly larger than those of RF for synthetic data 2 but smaller for T3 of DLBCL.

\begin{table}[!h]
\centering
\caption{Brier score for risk 1 of synthetic data 1.} 
\label{tab:linear_risk1}
\makebox[\linewidth]{
\resizebox{\linewidth}{!}{
\begin{tabular}{lcccccc}
 \hline
 & $\tau=0.5$ & $\tau=1$ & $\tau=1.5$ & $\tau=2$ & $\tau=2.5$ & $\tau=3$ \\ 
 \hline
Cox & .165$\pm$.012 & {\bf .166}$\pm$.010 & .165$\pm$.010 & .166$\pm$.012 & {\bf .164}$\pm$.012 & {\bf .162}$\pm$.012 \\ 
 FG & .168$\pm$.010 & .167$\pm$.010 & .166$\pm$.009 & .166$\pm$.012 & {\bf .164}$\pm$.013 & {\bf .162}$\pm$.012 \\ 
 BST & .167$\pm$.010 & {\bf .166}$\pm$.010 & .166$\pm$.010 & .166$\pm$.010 & .166$\pm$.011 & .165$\pm$.010 \\ 
 RF & .173$\pm$.013 & .175$\pm$.012 & .171$\pm$.012 & .172$\pm$.014 & .172$\pm$.014 & .170$\pm$.014 \\ 
 LDR & {\bf .164}$\pm$.014 & {\bf .166}$\pm$.011 & {\bf .164}$\pm$.010 & {\bf .165}$\pm$.012 & {\bf .164}$\pm$.013 & {\bf .162}$\pm$.013 \\ 
 \hline
\end{tabular}
}}
\end{table}

\begin{table}[!h]
\centering
\caption{Brier score for risk 2 of synthetic data 1.} 
\label{tab:linear_risk2}
\makebox[\linewidth]{
\resizebox{\linewidth}{!}{
\begin{tabular}{rllllll}
 \hline
 & $\tau= 0.5$ & $\tau= 1$ & $\tau= 1.5$ & $\tau= 2$ & $\tau= 2.5$ & $\tau= 3$ \\ 
 \hline
Cox & {\bf .152}$\pm$.011 & {\bf .158}$\pm$.014 & {\bf .158}$\pm$.015 & .157$\pm$.015 & {\bf .157}$\pm$.014 & .159$\pm$.014 \\ 
 FG & .157$\pm$.012 & .159$\pm$.014 & .159$\pm$.015 & .158$\pm$.015 & .158$\pm$.014 & .159$\pm$.014 \\ 
 BST & .158$\pm$.013 & {\bf .158}$\pm$.013 & {\bf .158}$\pm$.013 & .158$\pm$.013 & .158$\pm$.013 & .158$\pm$.013 \\ 
 RF & .164$\pm$.012 & .166$\pm$.015 & .166$\pm$.016 & .164$\pm$.015 & .165$\pm$.014 & .165$\pm$.014 \\ 
 LDR & {\bf .152}$\pm$.012 & {\bf .158}$\pm$.014 & {\bf .158}$\pm$.016 & {\bf .156}$\pm$.015 & {\bf .157}$\pm$.014 & {\bf .158}$\pm$.014 \\ 
 \hline
\end{tabular}
}}
\end{table}

\begin{table}[!h]
\centering
\caption{Brier score for risk 1 of synthetic data 2.} 
\label{tab:cosh_risk1}
\makebox[\linewidth]{
\resizebox{\linewidth}{!}{
\begin{tabular}{lcccccc}
 \hline
 & $\tau=1$ & $\tau=2$ & $\tau=3$ & $\tau=4$ & $\tau=5$ & $\tau=6$ \\ 
 \hline
Cox & .206$\pm$.008 & .235$\pm$.006 & .241$\pm$.005 & .242$\pm$.005 & .243$\pm$.005 & .243$\pm$.005 \\ 
 FG & .206$\pm$.008 & .235$\pm$.006 & .241$\pm$.006 & .242$\pm$.005 & .243$\pm$.005 & .243$\pm$.005 \\ 
 BST & .234$\pm$.005 & .234$\pm$.005 & .234$\pm$.005 & .234$\pm$.005 & .234$\pm$.005 & .234$\pm$.005 \\ 
 RF & {\bf .186}$\pm$.010 & {\bf .193}$\pm$.011 & {\bf .188}$\pm$.011 & {\bf .186}$\pm$.010 & {\bf .184}$\pm$.010 & {\bf .183}$\pm$.010 \\ 
 LDR & .193$\pm$.007 & .194$\pm$.007 & .191$\pm$.006 & .191$\pm$.006 & .191$\pm$.006 & .191$\pm$.006 \\ 
 \hline
\end{tabular}
}}
\end{table}

\begin{table}[!h]
\centering
\caption{Brier score for risk 2 of synthetic data 2.} 
\label{tab:cosh_risk2}
\makebox[\linewidth]{
\resizebox{\linewidth}{!}{
\begin{tabular}{lcccccc}
 \hline
 & $\tau=1$ & $\tau=2$ & $\tau=3$ & $\tau=4$ & $\tau=5$ & $\tau=6$ \\ 
 \hline
Cox & .251$\pm$.002 & .247$\pm$.003 & .245$\pm$.004 & .244$\pm$.004 & .244$\pm$.004 & .244$\pm$.004 \\ 
 FG & .251$\pm$.002 & .247$\pm$.003 & .245$\pm$.004 & .244$\pm$.004 & .244$\pm$.004 & .244$\pm$.005 \\ 
 BST & .245$\pm$.003 & .245$\pm$.003 & .245$\pm$.003 & .245$\pm$.003 & .245$\pm$.003 & .245$\pm$.003 \\ 
 RF & {\bf .178}$\pm$.011 & {\bf .182}$\pm$.010 & {\bf .181}$\pm$.010 & {\bf .182}$\pm$.010 & {\bf .182}$\pm$.010 & {\bf .183}$\pm$.010 \\ 
 LDR & .204$\pm$.006 & .199$\pm$.005 & .197$\pm$.005 & .198$\pm$.005 & .197$\pm$.005 & .199$\pm$.005 \\ 
 \hline
\end{tabular}
}}
\end{table}

\begin{table}[!h]
\centering
\caption{Brier score for ABC of DLBCL.} 
\label{tab:DLBCL_risk1}
\makebox[\linewidth]{
\resizebox{\linewidth}{!}{
\begin{tabular}{lcccccc}
 \hline
 & $\tau=1$ & $\tau=2$ & $\tau=3$ & $\tau=4$ & $\tau=5$ & $\tau=6$ \\ 
 \hline
Cox & .162$\pm$.056 & .190$\pm$.055 & .196$\pm$.058 & .202$\pm$.054 & .196$\pm$.053 & .202$\pm$.054 \\ 
 FG & .159$\pm$.057 & .185$\pm$.058 & .198$\pm$.058 & .196$\pm$.057 & .196$\pm$.056 & .199$\pm$.055 \\ 
 BST & { .136}$\pm$.045 & .146$\pm$.045 & {\bf .163}$\pm$.044 & {\bf .154}$\pm$.044 & {\bf .150}$\pm$.045 & \bf{.152}$\pm$.044 \\ 
 RF & .156$\pm$.052 & .173$\pm$.055 & .196$\pm$.051 & .198$\pm$.051 & .198$\pm$.051 & .200$\pm$.051 \\ 
 LDR & {\bf .131}$\pm$.050 & {\bf .143}$\pm$.050 & {\bf .163}$\pm$.047 & { .158}$\pm$.045 & { .155}$\pm$.043 & { .156}$\pm$.041 \\ 
 \hline
\end{tabular}
}}
\end{table}

\begin{table}[!h]
\centering
\caption{Brier score for GCB of DLBCL.} 
\label{tab:DLBCL_risk2}
\makebox[\linewidth]{
\resizebox{\linewidth}{!}{
\begin{tabular}{lcccccc}
 \hline
 & $\tau=1$ & $\tau=2$ & $\tau=3$ & $\tau=4$ & $\tau=5$ & $\tau=6$ \\ 
 \hline
Cox & .138$\pm$.048 & .212$\pm$.051 & .266$\pm$.061 & 268$\pm$.062 & .265$\pm$.062 & .277$\pm$.063 \\ 
 FG & .137$\pm$.046 & .206$\pm$.064 & .268$\pm$.059 & .265$\pm$.062 & .267$\pm$.063 & .273$\pm$.064 \\ 
 BST & .133$\pm$.046 & .204$\pm$.056 & .262$\pm$.042 & .252$\pm$.036 & .253$\pm$.048 & .257$\pm$.041 \\ 
 RF & .137$\pm$.038 & .197$\pm$.054 & .248$\pm$.050 & .247$\pm$.046 & .253$\pm$.050 & .262$\pm$.053 \\ 
 LDR & {\bf .129}$\pm$.035 & {\bf .179}$\pm$.052 & {\bf .242}$\pm$.053 & {\bf .236}$\pm$.050 & {\bf .237}$\pm$.052 & {\bf .244}$\pm$.052 \\ 
 \hline
\end{tabular}
}}
\end{table}

\begin{table}[!h]
\centering
\caption{Brier score for T3 of DLBCL.} 
\label{tab:DLBCL_risk3}
\makebox[\linewidth]{
\resizebox{\linewidth}{!}{
\begin{tabular}{lcccccc}
 \hline
 & $\tau=1$ & $\tau=2$ & $\tau=3$ & $\tau=4$ & $\tau=5$ & $\tau=6$ \\ 
 \hline
Cox & .193$\pm$.053 & .190$\pm$.061 & .206$\pm$.069 & .220$\pm$.071 & .233$\pm$.068 & .245$\pm$.072 \\ 
 FG & .183$\pm$.051 & .186$\pm$.062 & .195$\pm$.067 & .212$\pm$.069 & .230$\pm$.070 & .234$\pm$.069 \\ 
 BST & .169$\pm$.046 & .172$\pm$.044 & .177$\pm$.049 & .185$\pm$.046 & .185$\pm$.047 & .193$\pm$.048\\ 
 RF & .117$\pm$.045 & .151$\pm$.046 & .157$\pm$.043 & .169$\pm$.049 & .180$\pm$.051 & .185$\pm$.052 \\ 
 LDR & {\bf .111}$\pm$.035 & {\bf .137}$\pm$.038 & {\bf .142}$\pm$.036 & {\bf .151}$\pm$.041 & {\bf .165}$\pm$.044 & {\bf .171}$\pm$.046 \\ 
 \hline
\end{tabular}
}}
\end{table}

\begin{table}[!h]
\centering
\caption{Brier score for breast cancer of SEER.} 
\label{tab:seer_risk1}
\makebox[\linewidth]{
\resizebox{\linewidth}{!}{
\begin{tabular}{lccccccc}
 \hline
 & $\tau=10$ & $\tau=50$ & $\tau=100$ & $\tau=150$ & $\tau=200$ & $\tau=250$ & $\tau=300$ \\ 
 \hline
Cox & {\bf .014}$\pm$.003 & {\bf .106}$\pm$.006 & {\bf .150}$\pm$.006 & .169$\pm$.006 & .177$\pm$.007 & {\bf .180}$\pm$.006 & {\bf .179}$\pm$.005 \\ 
 FG & .016$\pm$.003 & .112$\pm$.011 & .156$\pm$.009 & .170$\pm$.006 & .177$\pm$.011 & .186$\pm$.013 & .189$\pm$.010 \\ 
 BST & {\bf .014}$\pm$.004 & .114$\pm$.008 & .154$\pm$.007 & {\bf .168}$\pm$.004 & {\bf .174}$\pm$.009 & .184$\pm$.009 & .184$\pm$.008 \\ 
 RF & .015$\pm$.003 & {\bf .106}$\pm$.007 & .151$\pm$.007 & .174$\pm$.008 & .182$\pm$.008 & .185$\pm$.008 & .187$\pm$.007 \\ 
 LDR & .018$\pm$.003 & .107$\pm$.006 & .153$\pm$.006 & .173$\pm$.006 & .182$\pm$.007 & .186$\pm$.006 & .185$\pm$.006 \\ 
 \hline
\end{tabular}
}}
\end{table}

\begin{table}[!h]
\centering
\caption{Brier score for other causes of SEER.} 
\label{tab:seer_risk2}
\makebox[\linewidth]{
\resizebox{\linewidth}{!}{
\begin{tabular}{rlllllll}
 \hline
 & $\tau=10$ & $\tau=50$ & $\tau=100$ & $\tau=150$ & $\tau=200$ & $\tau=250$ & $\tau=300$ \\ 
 \hline
Cox & {\bf .008}$\pm$.003 & {\bf .073}$\pm$.011 & {\bf .141}$\pm$.010 & .195$\pm$.010 & {\bf .204}$\pm$.010 & {\bf .193}$\pm$.009 & {\bf .178}$\pm$.007 \\ 
 FG & {\bf .008}$\pm$.003 & .076$\pm$.010 & .161$\pm$.013 & .241$\pm$.018 & .290$\pm$.029 & .302$\pm$.035 & .301$\pm$.040 \\ 
 BST & {\bf .008}$\pm$.003 & .074$\pm$.009 & .142$\pm$.011 & .201$\pm$.010 & .213$\pm$.016 & .203$\pm$.006 & .228$\pm$.018 \\ 
 RF & {\bf .008}$\pm$.003 & {\bf .073}$\pm$.010 & .145$\pm$.011 & .200$\pm$.010 & .213$\pm$.009 & .207$\pm$.009 & .199$\pm$.008 \\ 
 LDR & .009$\pm$.003 & .083$\pm$.008 & .148$\pm$.008 & {\bf .193}$\pm$.008 & .205$\pm$.009 & .199$\pm$.008 & .194$\pm$.008 \\ 
 \hline
\end{tabular}
}}
\end{table}

\begin{figure}[!ht]
 \centering
 \begin{subfigure}[t]{0.5\textwidth}
 \centering
\includegraphics[width=1\linewidth]{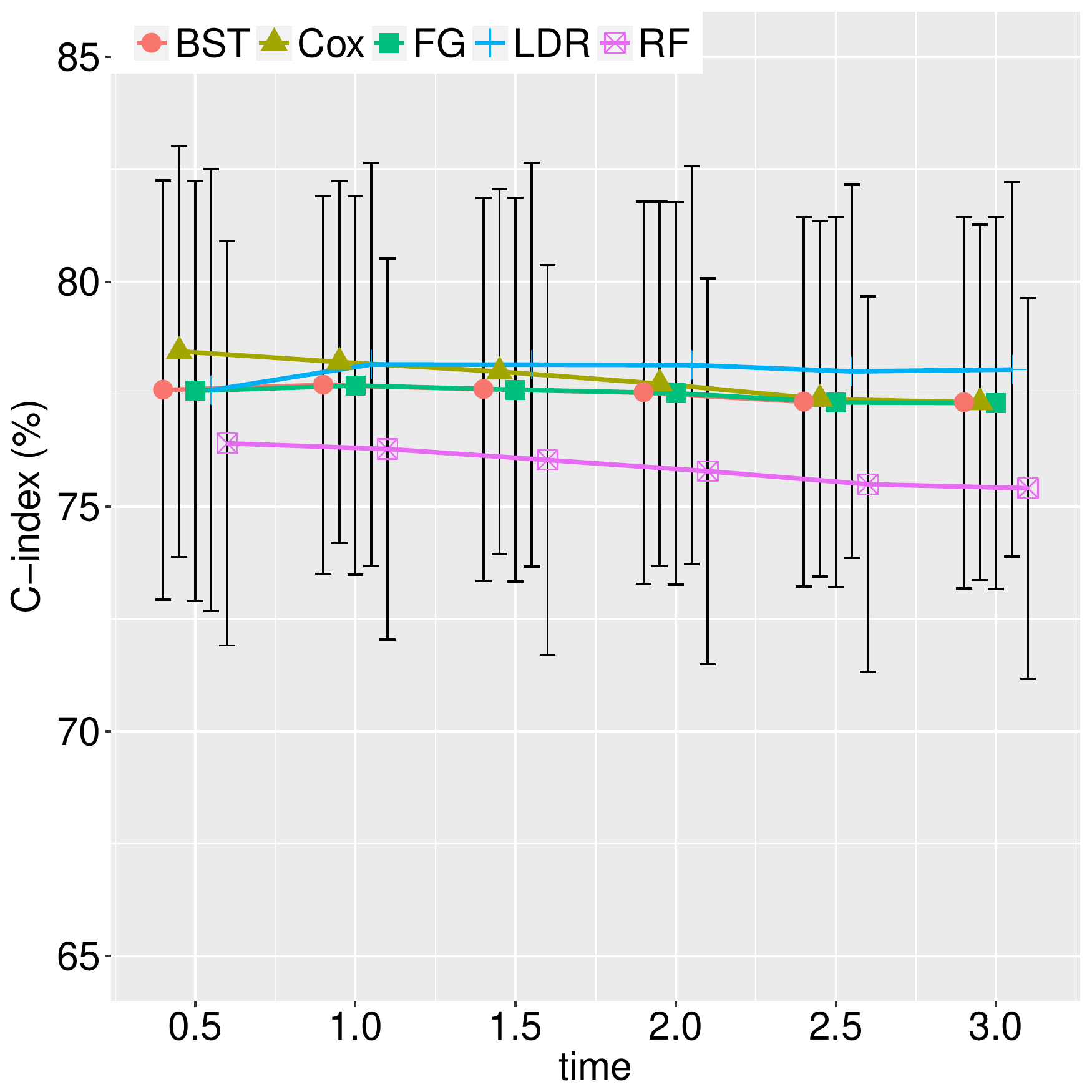}\vspace{-1.5mm}\label{linear_risk2}
 \caption{C-index of risk 2 for synthetic data 1.}\vspace{-1mm}
 \end{subfigure}%
~
\begin{subfigure}[t]{0.5\textwidth}
 \centering
\includegraphics[width=1\linewidth]{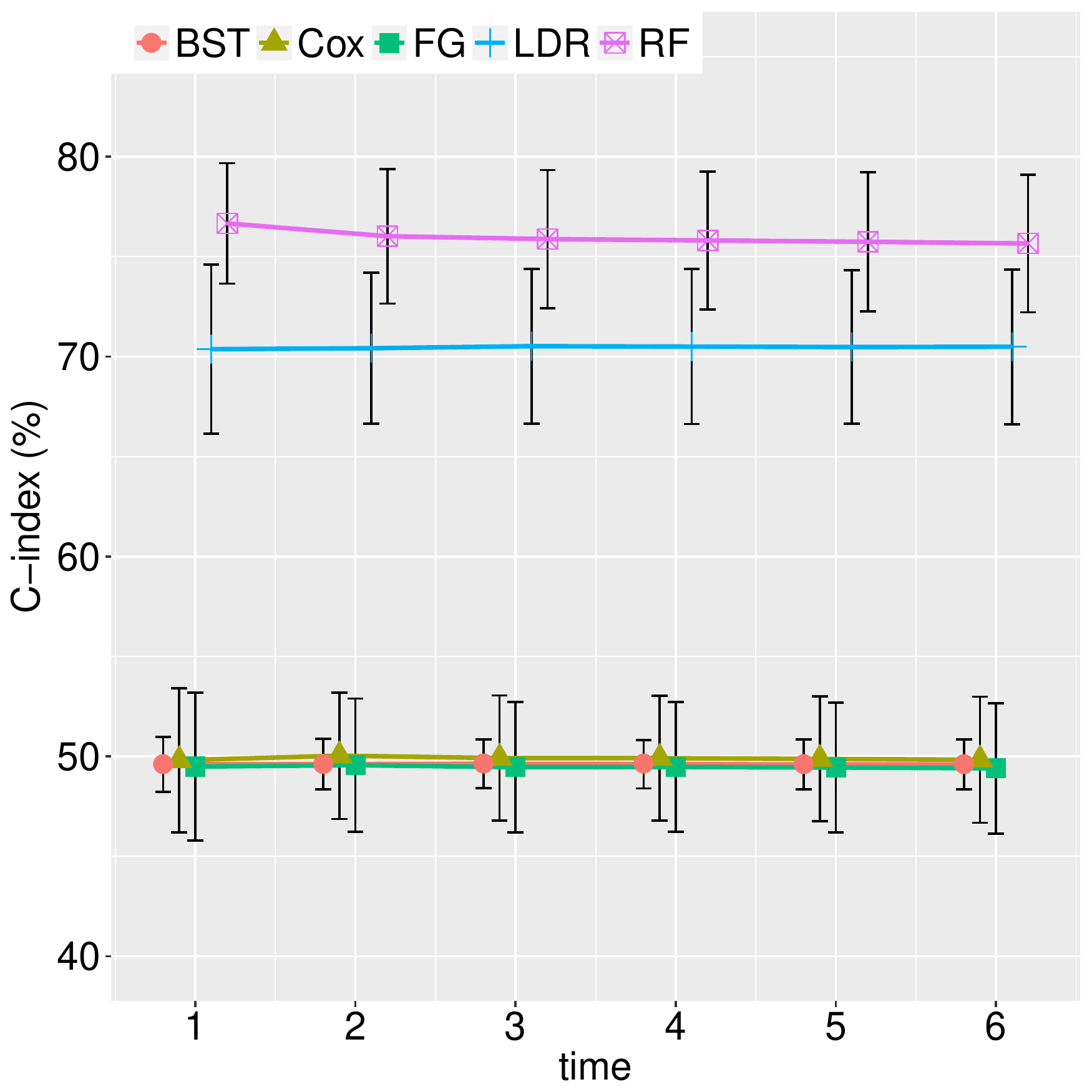}\vspace{-1.5mm}\label{cosh_risk1}
 \caption{C-index of risk 2 for synthetic data 2.}\vspace{-1mm}
 \end{subfigure}%
\caption{Cause-specific C-indices of risk 2 for synthetic data 1 and 2.
 }\vspace{-3mm}\label{toyrisk2}
\end{figure}

We show in Figure \ref{toyrisk2} the C-indices of risk 2 for synthetic data 1 and 2 used in Section \ref{sec:synthetic}. The C-indices of risk 2 for data 1 are very similar to those of risk 1 as in panel (a) of Figure \ref{linear_toy}; LDR, Cox, FG and BST are comparable and all slightly outperform RF in terms of mean values. The C-indices of risk 2 for data 2 are also analogous to those of risk 1 as in panel (c) of Figure \ref{linear_toy} except that LDR slightly underperforms RF in terms of mean values. But they both significally outperforms the other three approaches which completely fail.

Since we have random partitions in the analysis of DLBCL dataset, improvements of LDR can be underrated for the overlaps of boxplots across the five approaches in Figure \ref{lymphoma_cindex}. Therefore, we calculate the difference of C-indices between LDR and each of the other four benchmarks within each random partition, and report the mean and standard deviation in Table \ref{tab:lymphoma} where $\Delta_{\mbox{X}}$, $\mbox{X}\in\{\mbox{Cox},\mbox{ FG},\mbox{ BST},\mbox{ RF}\}$, denotes the C-index of LDR minus that of approach X. In terms of mean difference, LDR outperforms all the other benchmarks for all the three risks at any time evaluated except for BST under risk ABC. 

\begin{table}[!th]
\centering
\caption{Difference of C-indices between LDR and other benchmarks.} 
\label{tab:lymphoma}
\makebox[\linewidth]{
\resizebox{\linewidth}{!}{
\begin{tabular}{l|cccc|cccc|cccc}
 \toprule
 & \multicolumn{4}{c}{ABC} & \multicolumn{4}{c}{GCB} & \multicolumn{4}{c}{T3} \\
year & $\Delta_{\mbox{COX}}$ & $\Delta_{\mbox{FG}}$ & $\Delta_{\mbox{BST}}$ & $\Delta_{\mbox{RF}}$ & $\Delta_{\mbox{COX}}$ & $\Delta_{\mbox{FG}}$ & $\Delta_{\mbox{BST}}$ & $\Delta_{\mbox{RF}}$ & $\Delta_{\mbox{COX}}$ & $\Delta_{\mbox{FG}}$ & $\Delta_{\mbox{BST}}$ & $\Delta_{\mbox{RF}}$ \\ 
 \midrule
1 & .09$\pm$.08 & .03$\pm$.05 & .01$\pm$.06 & .06$\pm$.08 & .07$\pm$.09 & .06$\pm$.09 & .07$\pm$.06 & .16$\pm$.12 & .16$\pm$.15 & .11$\pm$.12 & .06$\pm$.05 & .10$\pm$.12 \\ 
 2 & .09$\pm$.06 & .03$\pm$.04 & .00$\pm$.07 & .04$\pm$.08 & .11$\pm$.08 & .10$\pm$.08 & .05$\pm$.06 & .17$\pm$.13 & .20$\pm$.17 & .10$\pm$.08 & .05$\pm$.05 & .03$\pm$.08 \\ 
 3 & .09$\pm$.05 & .04$\pm$.05 & -.01$\pm$.06 & .05$\pm$.06 & .12$\pm$.07 & .12$\pm$.06 & .05$\pm$.06 & .16$\pm$.09 & .20$\pm$.17 & .10$\pm$.09 & .05$\pm$.05 & .03$\pm$.08 \\ 
 4 & .09$\pm$.05 & .04$\pm$.05 & -.01$\pm$.06 & .05$\pm$.06 & .11$\pm$.07 & .12$\pm$.06 & .05$\pm$.06 & .15$\pm$.10 & .21$\pm$.15 & .11$\pm$.09 & .04$\pm$.05 & .02$\pm$.08 \\ 
 5 & .09$\pm$.05 & .04$\pm$.05 & -.01$\pm$.06 & .05$\pm$.06 & .12$\pm$.07 & .12$\pm$.06 & .05$\pm$.06 & .15$\pm$.09 & .21$\pm$.16 & .11$\pm$.08 & .04$\pm$.05 & .01$\pm$.08 \\ 
 6 & .09$\pm$.05 & .03$\pm$.05 & -.01$\pm$.06 & .04$\pm$.06 & .11$\pm$.07 & .12$\pm$.06 & .05$\pm$.06 & .16$\pm$.09 & .23$\pm$.14 & .11$\pm$.08 & .04$\pm$.05 & .02$\pm$.09 \\ 
 \bottomrule
\end{tabular}
}}
\end{table}

\end{document}